%% file: main.tex
\pdfoutput=1
\documentclass[sigconf,screen]{acmart}
\usepackage[utf8]{inputenc}

\input{math_commands.tex}

\usepackage{microtype} %

\usepackage[inkscapeopt={-D}]{svg}

\usepackage{url}
\usepackage{graphicx} 

\usepackage{newtxtext,newtxmath}
\usepackage{xcolor}
\usepackage{subcaption}
\usepackage{placeins}
\usepackage[most]{tcolorbox}
\usepackage{adjustbox}
\usepackage{colortbl}
\usepackage{ifthen}
\usepackage{multirow}
\usepackage{booktabs}
\usepackage{threeparttable}
\usepackage[detect-weight]{siunitx}
\usepackage{pifont}   %

\newcommand{\PCT}{\%}

\usepackage[ruled,vlined]{algorithm2e}

\definecolor{rqoneink}{HTML}{1F6F5C}\definecolor{rqonebar}{HTML}{44AA99}   %
\definecolor{rqtwoink}{HTML}{332288}\definecolor{rqtwobar}{HTML}{6A5FBF}   %
\definecolor{rqthreeink}{HTML}{882255}\definecolor{rqthreebar}{HTML}{CC6677} %
\colorlet{rqink}{rqoneink}\colorlet{rqbar}{rqonebar}  %
\newcommand{\setrq}[1]{\colorlet{rqink}{rq#1ink}\colorlet{rqbar}{rq#1bar}}
\newtcolorbox{finding}{
  enhanced, breakable, rounded corners, arc=2.5pt,
  colback=black!5, colframe=black!5, boxrule=0pt,
  left=8pt, right=6pt, top=4pt, bottom=4pt,
  before skip=6pt, after skip=6pt, fontupper=\small,
  overlay unbroken and first={
    \begin{scope}
      \clip[rounded corners=2.5pt]
        (frame.south west) rectangle (frame.north east);
      \fill[rqbar]
        (frame.south west) rectangle ([xshift=2.6pt]frame.north west);
    \end{scope}
  },
}
\newcommand{\findhead}[1]{\textbf{\textcolor{rqink}{#1.}}\enspace}
\newcommand{\rqlabel}[2]{\textcolor{rq#1ink}{\textbf{#2}}}
\newcommand{\rqsection}[2]{\par\textcolor{rq#1ink}{\textbf{#2}\;}}
\newcommand{\rqref}[2]{\textcolor{rq#1ink}{\textit{#2}}}  %

\newboolean{debugmode}
\setboolean{debugmode}{false}  %

\newcommand{\includesvgwithadjust}[4][]{%
  \ifthenelse{\boolean{debugmode} \AND\NOT\equal{#1}{nodebug}}{%
    \setlength{\fboxsep}{0pt}%
    \setlength{\fboxrule}{0.5pt}%
    \color{red}%
    \fbox{%
      \color{black}%
      \adjustbox{trim=#3}{%
        \includesvg[width=#2\linewidth]{#4}%
      }%
    }%
  }{%
    \adjustbox{trim=#3}{%
      \includesvg[width=#2\linewidth]{#4}%
    }%
  }%
}

\newcommand{\ccpp}{\texttt{C/C++}}
\newcommand{\cve}{\texttt{CVE}}
\newcommand{\cves}{\texttt{CVEs}}

\title{A Comprehensive Evaluation of Code Language Models for Security Patch Detection}
\author{Nils Loose}
\correspondingauthor
\orcid{0009-0003-6243-1623}
\affiliation{%
  \institution{University of Lübeck}
  \city{Lübeck}
  \country{Germany}}
\email{n.loose@uni-luebeck.de}
\author{Joseph Bienhüls}
\orcid{0009-0005-5016-354X}
\affiliation{%
  \institution{University of Lübeck}
  \city{Lübeck}
  \country{Germany}}
\email{j.bienhuels@uni-luebeck.de}
\author{Kristoffer Hempel}
\orcid{0009-0008-9159-4268}
\affiliation{%
  \institution{University of Lübeck}
  \city{Lübeck}
  \country{Germany}}
\email{k.hempel@uni-luebeck.de}
\author{Felix Mächtle}
\orcid{0009-0009-2431-0322}
\affiliation{%
  \institution{University of Lübeck}
  \city{Lübeck}
  \country{Germany}}
\email{f.maechtle@uni-luebeck.de}
\author{\mbox{Thomas Eisenbarth}}
\orcid{0000-0003-1116-6973}
\affiliation{%
  \institution{University of Lübeck}
  \city{Lübeck}
  \country{Germany}}
\email{thomas.eisenbarth@\allowbreak{}uni-luebeck.de}
\renewcommand{\authors}{Nils Loose\and Joseph Bienhüls\and Kristoffer Hempel\and Felix Mächtle\and Thomas Eisenbarth}

\newcommand{\axtext}[1]{\fontsize{8}{9}\selectfont #1}
\newcommand{\ticktext}[1]{\fontsize{7}{8}\selectfont #1}
\newcommand{\legendtext}[1]{\fontsize{7}{8}\selectfont #1}
\newcommand{\legt}[1]{\fontsize{7}{8}\selectfont #1}
\newcommand{\figtext}[1]{\fontsize{7}{8}\selectfont #1}

\newcommand{\ds}[1]{$\mathcal{D}_{#1}$}

\newcommand{\diff}{\texttt{diff}}
\newcommand{\diffs}{\texttt{diffs}}

\newcommand{\codebert}{CodeBERT}
\newcommand{\ctfive}{CodeT5-Large}
\newcommand{\vfc}{\texttt{VFC}}
\newcommand{\vfcs}{\texttt{VFCs}}
\newcommand{\spd}{\texttt{SPD}}
\newcommand{\vulndet}{\texttt{VD}}
\newcommand{\ml}{\texttt{ML}}

\copyrightyear{2026}
\acmYear{2026}
\setcopyright{cc}
\setcctype{by}
\acmConference[ASE '26]{Proceedings of the 41st IEEE/ACM International Conference on Automated Software Engineering}{October 12--16, 2026}{Munich, Germany}
\acmBooktitle{Proceedings of the 41st IEEE/ACM International Conference on Automated Software Engineering (ASE '26), October 12--16, 2026, Munich, Germany}
\acmDOI{10.1145/3832783.3837538}
\acmISBN{979-8-4007-2882-2/2026/10}
\acmSubmissionID{ase26main-p2677-p}
\begin{document}

\begin{abstract}
Automated detection of vulnerability-fixing commits (\vfcs) is critical for timely security patch deployment, as advisory databases lag patch releases by a median of 25 days and many fixes never receive advisories. Code language models are increasingly adopted for identifying \vfcs, yet whether they can recognize a security fix from the code change itself remains unclear, as reported performance is shaped by commit messages, project-level data leakage, and uncertain data quality. We present a rigorous re-evaluation that jointly controls these factors through a unified framework consolidating \num{20} fragmented datasets spanning more than \num{180000} commits. Training \num{270} models from \num{125}M to \num{80}B parameters, we isolate the code signal under group-stratified code-only evaluation and assess the impact of model capacity and additional code context, ranging from intra-procedural enrichment to inter-procedural repository context. Model capacity yields clear but insufficient gains, and the evaluated context signals provide no reliable improvement under strict false positive budgets. At a false positive rate of \num{0.5}\%, every evaluated fine-tuned code-only model misses at least \num{80}\% of vulnerability fixes. A manual expert audit further shows that label error concentrates in commits lacking \cve\ association and primarily distorts evaluation. We derive concrete recommendations for evaluating on aggregated \vfc\ datasets and release our unified framework and evaluation suite.
\end{abstract}



\maketitle

\section{Introduction}\label{introduction}
\input{figures/rl-vfc-datasets.tex}

\input{sections/Introduction.tex}

\section{Related Work}\label{related-work}\FloatBarrier
\input{sections/RelatedWork.tex}
\input{figures/framework/datasets.tex}\section{\vfc\ Collection Framework}\label{vfc-collection-framework}
\input{sections/Framework.tex}
\section{Evaluation}\label{evaluation}
\input{sections/Evaluation.tex}
\section{Conclusion}\label{conclusion}
\input{sections/Conclusion.tex}

\begin{acks}
This work has been supported by funding from the Agentur für Innovation in der Cybersicherheit GmbH (Cyberagentur, project SOVEREIGN).
\end{acks}

\appendix
\section{Context Enrichment Algorithm}\label{app:context-alg}
\input{figures/context-alg.tex}

\FloatBarrier

\section*{Data Availability}
\sloppy
We release our framework at \url{https://github.com/UzL-ITS/vfc_datasets} together with a comprehensive artifact archive~\citep{vfcdetective_artifact} containing (1) the unified \vfc\ framework with automated download, normalization, deduplication, and splitting pipelines, (2) the context enrichment tool, (3) the full training and evaluation pipeline including preprocessing, context-aware truncation, split generation, and training configurations, and (4) complete training logs. As licensing restrictions prevent direct redistribution of the raw data and trained model weights, the framework instead automates the full data generation once researchers obtain access to the gated source collections.

\bibliographystyle{ACM-Reference-Format}
\bibliography{main}
\end{document}

%% file: math_commands.tex
\usepackage{amsmath,amsfonts,bm}

\def\eqref#1{equation~\ref{#1}}

\def\1{\bm{1}}

\DeclareMathAlphabet{\mathsfit}{\encodingdefault}{\sfdefault}{m}{sl}
\SetMathAlphabet{\mathsfit}{bold}{\encodingdefault}{\sfdefault}{bx}{n}

%% file: figures/rl-vfc-datasets.tex
\begin{figure*}[t]
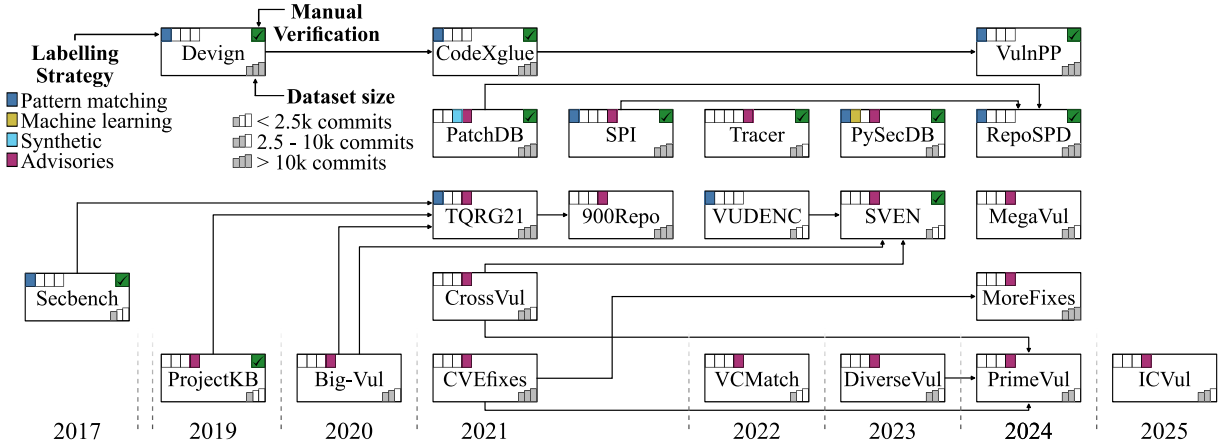

    \centering
    \includesvgwithadjust{0.9 }{0pt 0pt 0pt 0pt}{data/rl-temporal-overview.svg}
    \caption{Temporal overview of existing \vfc\ datasets, their label source, size, and dependent datasets. The advisory label class includes other, similar, sources of information such as bug trackers. The synthetic label indicates that at least a subset of samples is synthetic. Manual verification indicates that some effort was spent validating at least a subset of the samples. VulnPP: VulnPatchPairs.}\label{fig:rl-temporal-overview}
      \Description{Timeline diagram of vulnerability-fixing commit datasets from 2017 to 2025. Each dataset appears at its release year with markers encoding its label source, such as advisory linking, pattern matching, manual verification, or synthetic generation, and its sample count. Arrows connect datasets that build on earlier ones, showing that most later datasets reuse a small set of early sources.}
  \end{figure*}

%% file: sections/Introduction.tex
Modern software supply chains create cascading security dependencies where vulnerabilities impact numerous downstream projects. Organizations must identify and apply patches before attackers exploit the disclosure window. However, \citet{DBLP:journals/tse/ImtiazKW23} reveal that advisory publication lags patch release by a median of 25 days, leaving systems vulnerable to attacks when developers rely on vulnerability databases such as the National Vulnerability Database (NVD)~\citep{nvd}. Silent security patches present an even greater challenge, as they never receive public advisories, leaving downstream software vulnerable. As the vast majority of reused software is maintained in open-source repositories, an accurate analysis of each commit across upstream dependencies could result in safer software ecosystems with faster patch deployment times. However, as manual analysis is infeasible at scale, automated techniques for identifying vulnerability-fixing commits (\vfcs) have become essential. More broadly, as code changes in diff format have become the standard unit of review in modern development, from manual code review to AI-assisted programming, understanding the reasoning capabilities of current models on software changes is relevant well beyond the security domain.
Code LMs~\citep{codebert,unixcoder,CodeT52021} have shown promise across code understanding tasks, leading to their adoption for Security Patch Detection (\spd), the task of identifying commits that fix vulnerabilities. Existing approaches report strong results when classifying full commits~\citep{DBLP:journals/corr/abs-2308-15233,DBLP:conf/milcom/WangWFSJBG21,10.1145/3749370,DBLP:conf/icsm/SabettaB18,DBLP:conf/wcre/Nguyen-TruongKL22, CompVPD:chen2024compvpditerativelyidentifyingvulnerability, DBLP:conf/icse/SunXLXZHZ23}. However, three properties of this setting obscure what such results reveal about code understanding. Ablation studies attribute this performance largely to the commit message rather than the code changes~\citep{CompVPD:chen2024compvpditerativelyidentifyingvulnerability,DBLP:conf/icse/SunXLXZHZ23}. \vfc\ corpora cluster by project, and evaluations that share projects across splits have been shown to reward memorization rather than generalization on vulnerability detection (\vulndet)~\citep{DBLP:conf/icse/SteenhoekRJL23,DBLP:conf/uss/ArpQPWPWCR22}. Finally, label quality is a known concern for \vfc\ data, yet error rates have been quantified primarily for the related task of function-level vulnerability detection~\citep{PrimeVul,DiverseVul}. Each factor has been observed in isolation, but their joint effect on reported \spd\ performance remains unquantified. In this work, we perform a rigorous re-evaluation of code LM-based \spd\ under an evaluation protocol that jointly addresses all three factors.

To enable this re-evaluation at scale, we first address the data fragmentation challenge. Over the past decade, a variety of research groups have collected and labeled \vfcs\ with different target languages, labeling techniques, and ground truths, resulting in fragmented datasets that hinder large-scale comparative evaluation. We build a comprehensive framework\footnote{\url{https://github.com/UzL-ITS/vfc_datasets}} that unifies access to \num{20} \vfc\ datasets, enabling systematic evaluation across different labeling strategies, programming languages, and vulnerability classes. To evaluate whether additional semantic context can improve model behavior, we additionally implement a lightweight intra-procedural context enrichment method that identifies semantically relevant code context around modifications.

We conduct a comprehensive re-evaluation of code LM-based \spd, training \num{270} distinct models across architectures from \num{125}M dense to \num{80}B mixture-of-experts parameters, multiple input representations, and splitting strategies, requiring over \num{2200} GPU hours. Confirming prior observations at scale, we find that the evaluated models classify commits primarily through the message, with message-only training nearly matching full-commit performance. In the code-only setting, group-stratified splits that separate repository groups between partitions reduce F1 by up to \num{28}\% compared to random splits, revealing project-specific memorization. Temporal splits degrade further, but a sliding window analysis attributes this to shifting project compositions rather than genuine temporal shift. Faithful temporal evaluation therefore requires a dataset composition that is stable over time, which current aggregated \vfc\ corpora do not provide.

Under this protocol, neither the evaluated model capacity nor the enriched context signals resolve the task. Encoder-based models barely exceed a lexical TF-IDF baseline, scaling to \num{80}B parameters yields clear but insufficient gains, and enriching inputs with control- and data-flow-based intra-procedural context or inter-procedural repository context~\citep{RepoSPD} does not improve detection consistently. Our manual audit further shows that label error concentrates in \vfcs\ lacking \cve\ association and primarily distorts evaluation, while in our cross-dataset experiment, \cve-mapped \vfcs\ outweigh sheer training data volume under strict false positive budgets. At a false positive rate of \num{0.5}\%, every fine-tuned code-only configuration misses at least \num{80}\% of vulnerability fixes. Our error analysis suggests that the decisive evidence for a fix often lies outside the \diff, while models fail to benefit even from semantically relevant added context. Designing systems that identify and effectively use such context therefore emerges as a central open problem for \spd.

In summary, we make three contributions. (1) We build a unified framework consolidating \num{20} fragmented \vfc\ datasets alongside lightweight context enrichment tooling~\citep{vfcdetective_artifact}. (2) We rigorously re-evaluate code LM-based \spd\ along three research questions covering the training and evaluation protocol, model capacity and context, and label and dataset quality, deriving concrete recommendations for evaluating on aggregated \vfc\ corpora. (3) We directly quantify commit-level label accuracy through an expert audit and a systematic error analysis, isolating how label noise affects training and evaluation.

%% file: sections/RelatedWork.tex
\input{tables/datasets.tex}\par\textbf{\vfc\ Datasets}\label{vfc-datasets}
Security research focusing on \spd\ or \vulndet\ requires code samples that are annotated with accurate vulnerability information. To approximate the underlying structural distribution of the data, a sufficiently large sample size is required. Yet, obtaining positive ground truth samples is difficult. Over the past decade, an increasingly popular method of obtaining security-annotated samples revolves around \vfcs.
To obtain code samples, open-source repositories hosted on public version control systems are the primary source. The process of identifying which commit is security relevant has seen three major approaches: (i)~algorithmic pattern matching, often on the commit message~\citep{Devign,SecBenchReis,tqrg,SPI,VUDENC,PySecDB}, (ii)~linking vulnerability information from public advisories, bug trackers, or similar sources to commits~\citep{Vulas,BigVul,CVEfixes,CrossVul,tqrg,PatchDB,SPI,900repo,tracer,PySecDB,MegaVul,MoreFixes,PrimeVul,ICVul}, and (iii)~using other tools, algorithms or machine learning (\ml) to perform code-based identification of potential \vfcs~\citep{PatchDB,PySecDB,VulnPatchPairs,SPI,Sven,Vulas}.
Additionally, synthetic generation of \vfcs\ has been explored~\citep{PatchDB}. Regardless of the technique used for generating labels, it has been shown~\citep{PrimeVul} that the quality of labels is a significant concern. Some works have attempted to quantify this by using manual verification~\citep{SecBenchReis,Devign,Vulas,SPI,tracer,PySecDB}. An overview of dataset relations and key characteristics is shown in Figure~\ref{fig:rl-temporal-overview}. While some connections exist, the landscape remains fragmented, hindering large-scale comparative evaluations. An in-depth summary of the identified datasets, their size, label distribution, and included programming languages can be found in Table~\ref{tab:overview-datasets}.
\par\textbf{\vfc\ Detection}\label{vfc-detection}
Recent work has raised fundamental questions about what \ml\ models actually learn for security-related code understanding. \citet{VulnPatchPairs} show that top-performing vulnerability detection models overfit to label-unrelated features and cannot distinguish vulnerable functions from their patched versions. \citet{TopScoreWrongExam:Risse2025} survey 81 ML4VD papers and demonstrate that a simple word-count classifier achieves an F1 score comparable to that of deep learning models on popular benchmarks, suggesting that reported progress reflects dataset correlations rather than vulnerability understanding. These findings motivate a careful examination of \spd\ approaches.
Detecting \vfcs\ has been explored under several assumptions, from \spd, where the goal is to identify whether a commit is security relevant~\citep{GraphSPD,RepoSPD,SPI,CompVPD:chen2024compvpditerativelyidentifyingvulnerability,DBLP:conf/kbse/ZhouPW00WH21,DBLP:journals/pacmse/YangZPZWHL25,DBLP:conf/icse/SunXLXZHZ23}, to ranking-based approaches trying to identify the \vfc\ belonging to a specific advisory~\citep{VFCFinder,VCMatch}. Ranking-based approaches sometimes include a submodule that provides an \spd\ prediction indicator~\citep{VFCFinder}, embedded into a larger framework that ranks \vfcs\ using additional metadata. For \spd, several works have also explored the prediction capabilities under the utilization of the code changes along with the commit message~\citep{DBLP:journals/corr/abs-2308-15233,DBLP:conf/milcom/WangWFSJBG21,10.1145/3749370,DBLP:conf/icsm/SabettaB18,DBLP:conf/wcre/Nguyen-TruongKL22}.
In this work we evaluate \spd\ under the restriction that only the code is utilized for the classification to understand the security-related code reasoning capabilities.
Sequence-based approaches typically rely on representing code changes as token sequences that are used for training an \ml\ model or ensemble of models for supervised classification~\citep{DBLP:conf/icse/SunXLXZHZ23, VFCFinder, DBLP:conf/kbse/ZhouPW00WH21}. \citet{CoLeFunDA:DBLP:conf/icse/ZhouPCHXLH23} explore a contrastive learning approach to pretrain a BERT model with a specialized embedding space that is used in downstream \spd\ systems.
\citet{GraphSPD} introduced PatchCPG, a semantics-aware patch representation based on code property graphs~\citep{Joern}. To generate a PatchCPG, the CPG is generated for the pre- and post-patch versions using Joern~\citep{Joern}. Then, dependence-guided forward and backward slicing, seeded by the added and deleted lines, is performed to confine context. Finally, GraphSPD feeds a multi-attributed graph into a graph neural network (GNN) to classify security patches directly from graph structure~\citep{GraphSPD}. Building on this idea, \citet{RepoSPD} construct a RepoCPG that preserves semantic changes while augmenting them with cross-file dependencies (e.g., function-level call relations), and fuses graph-based and sequence-based representations with progressive learning to capture relationships among multiple code changes at repository scale~\citep{RepoSPD}. Recently, a dynamic approach to evaluating safe patches by integrating dynamic symbolic execution has also been proposed~\citep{DBLP:conf/icse/Luo0024}. Additionally, \citet{DBLP:journals/pacmse/YangZPZWHL25} propose an LLM-based generative approach that utilizes retrieval-augmented generation (RAG) through embedding-based matching of historical \vfc\ information to detect \vfcs.

%% file: tables/datasets.tex
\begin{table*}[t]
\begingroup
\small
\renewcommand{\arraystretch}{1.1}
\centering
\begin{threeparttable}
\caption{Overview of research-based datasets that contain \vfcs.}\label{tab:overview-datasets}
\begin{tabular}{llrrrrl@{\hskip 2pt}c@{\hskip 2pt}c@{\hskip 2pt}c@{\hskip 2pt}c}
\toprule
\multirow{2}[2]{*}{Year} & \multirow{2}[2]{*}{Dataset} & \multicolumn{3}{c}{\hspace*{-2em}Size} & \multirow{2}[2]{*}{\cves} & \multirow{2}[2]{*}{Languages} & \multicolumn{4}{c}{Included} \\
\cmidrule(lr{.5em}){3-5}
\cmidrule(lr{.5em}){8-11}
& & Repos & \textsc{VFCs} & \textsc{$\neg$VFCs} & & & \ds{1} & \ds{2} & \ds{3} & \ds{4} \\

\midrule
2017 & Secbench~\citep{SecBenchReis} & \num{112} & \num{639} & -- & \num{183} & \num{13} \texttt{PLs} & (\checkmark) & \checkmark & \checkmark & \checkmark\\
\midrule
2019 & Project KB~\citep{Vulas} & \num{189} & \num{1045} & -- & \num{551} & \texttt{Java}& \ding{55} & \ding{55} & \ding{55} & \checkmark\\
& Devign~\citep{Devign} & \num{2} & \num{11091} & \num{14587} & \num{386} & \ccpp & (\checkmark) & \checkmark & \checkmark & \checkmark\\
\midrule
2020 & Big-Vul~\citep{BigVul} & \num{413} & \num{4235} & -- & \num{3808} & \ccpp & (\checkmark) & \checkmark & \checkmark & \checkmark\\
\midrule
2021 & CVEfixes~\citep{CVEfixes} & \num{3925} & \num{11329} & -- & \num{11156} & \num{27} \texttt{PLs} & (\checkmark) & \checkmark & \checkmark & \checkmark\\
& SPI~\citep{SPI} & \num{2} & \num{11086} & \num{14573} & \num{386} & \ccpp & (\checkmark) & \checkmark & \checkmark & \checkmark\\
& PatchDB~\citep{PatchDB} & \num{391} & \num{9583} & \num{22806} & \num{2338} & \ccpp & (\checkmark) & \checkmark & \checkmark & \checkmark\\
& TQRG21~\citep{tqrg} & \num{1298} & \num{7627} & \num{85442} & \num{6079} & \num{20} \texttt{PLs} & (\checkmark) & \checkmark & \checkmark & \checkmark\\
& CrossVul~\citep{CrossVul} & \num{1518} & \num{5432} & -- & \num{5166} & \num{48} \texttt{PLs}\tnote{*} & (\checkmark) & \checkmark & \checkmark & \checkmark\\
& CodeXGLUE~\citep{DBLP:conf/nips/LuGRHSBCDJTLZSZ21}\tnote{†} & \num{4} & \num{10894} & -- & -- & \ccpp & \ding{55} & \ding{55} & \ding{55} & \ding{55} \\
& 900Repo~\citep{900repo} & \num{910} & \num{3189} & \num{5475} & \num{2545} & \num{20} \texttt{PLs} & (\checkmark) & \checkmark & \checkmark & \checkmark\\
\midrule
2022 & VCMatch~\citep{VCMatch} & \num{10} & \num{1568} & -- & \num{1631} & \ccpp, \texttt{Java}, \texttt{PHP} & (\checkmark) & \checkmark & \checkmark & \checkmark\\
& Tracer~\citep{tracer} & \num{727} & \num{2389} & -- & \num{1308} & \num{7}+ \texttt{PLs} & (\checkmark) & \checkmark & \checkmark & \checkmark\\
& VUDENC~\citep{VUDENC} & \num{784} & \num{1009} & -- & -- & \texttt{Python}& \ding{55} & \ding{55} & \ding{55} & \ding{55} \\
\midrule
2023 & DiverseVul~\citep{DiverseVul} & \num{797} & \num{7514} & -- & -- & \ccpp & \ding{55} & \checkmark & \checkmark & \checkmark\\
& PySecDB~\citep{PySecDB} & \num{476} & \num{668} & \num{1823} & \num{302} & \texttt{Python} & \ding{55} & \ding{55} & \ding{55} & \checkmark \\
& Sven~\citep{Sven} & \num{261} & \num{532} & -- & \num{417} & \ccpp, \texttt{Py} & (\checkmark) & \checkmark & \checkmark & \checkmark\\
\midrule
2024 & MoreFixes~\citep{MoreFixes} & \num{7037} & \num{32358} & -- & \num{26197} & \num{54} \texttt{PLs} & (\checkmark) & (\checkmark) & \checkmark & \checkmark\\
& MegaVul~\citep{MegaVul} & \num{1393} & \num{9108} & -- & \num{8575} & \ccpp, \texttt{Java} & (\checkmark) & \checkmark & \checkmark & \checkmark\\
& VulnPatchPairs~\citep{VulnPatchPairs} & 2 & \num{6352} & -- & -- & \ccpp & \ding{55} & \ding{55} & \ding{55} & \ding{55} \\
& PrimeVul~\citep{PrimeVul} & \num{755} & \num{6827} & -- & \num{5369} & \ccpp & \ding{55} & \ding{55} & \ding{55} & \ding{55} \\
& RepoSPD~\citep{RepoSPD} & \num{363} & \num{17341} & \num{30447} & \num{2323} & \ccpp & (\checkmark) & \checkmark & \checkmark & \checkmark\\
\midrule
2025 & ICVul~\citep{ICVul} & \num{658} & \num{3828} & \num{6} & \num{3730} & \ccpp & \checkmark & \checkmark & \checkmark & \checkmark\\
\bottomrule
\end{tabular}
\begin{tablenotes}
  \item[*] Unique file extensions.
  \item[†] Defect detection set is identical to Devign~\cite{Devign}, CodeXGLUE contains several other tasks
\end{tablenotes}
\end{threeparttable}
\endgroup
\end{table*}

%% file: figures/framework/datasets.tex
\begin{table}[h]
  \centering
  \caption{Evaluation dataset compositions. }
  \label{tab:dataset-overview}
  \setlength{\tabcolsep}{4pt}
  \begin{tabular}{c | c | l || r | r || r | r}
  \toprule
  \multirow{2}{*}{\textbf{Dataset}} & \multirow{2}{*}{\textbf{PLs}} & \multirow{2}{*}{\textbf{Label}} & \multicolumn{2}{c||}{\textbf{Commits}} & \multicolumn{2}{c}{\textbf{Projects}} \\
  & & & \textbf{Total} & \textbf{VFCs} & \textbf{100\%} & \textbf{75\%} \\
  \midrule
  \midrule
  \ds{1} & \ccpp & \cve & \num{35399} & \num{32}\% & \num{1144} & \num{49} \\
  \ds{2} & \ccpp & \texttt{Adv} & \num{88687} & \num{32}\% & \num{1341} & \num{35} \\
  \ds{3} & \ccpp & \texttt{All} & \num{96021} & \num{38}\% & \num{1698} & \num{46} \\
  \midrule
  \ds{4} & \texttt{Multi} & \texttt{All} & \num{183106} & \num{32}\% & \num{8092} & \num{314} \\
  \bottomrule
  \end{tabular}
\end{table}

%% file: sections/Framework.tex
\input{figures/dataset-characterizations.tex}The fragmented landscape of \vfc\ datasets (Section~\ref{vfc-datasets}) presents challenges for comparative evaluation, as accessibility issues, format inconsistencies, and varying labeling methodologies across individual collections make it difficult to compare results and reproduce findings across studies.
To address these challenges, we develop a framework that unifies access to existing \vfc\ datasets through systematic parsing, normalization, and enrichment. The framework consists of three stages: data ingestion and normalization across diverse formats and platforms, enrichment with commit metadata and repository information, and deduplication and filtering for dataset customization. The framework currently integrates 20 different data sources, enabling researchers to combine and filter collections based on specific research requirements.
To manage the overlap during dataset merging, we implement two complementary deduplication strategies. First, identical entries (same commit hash and repository) are merged to maximize metadata retention while removing conflicts where labels disagree. Second, semantic matches based on diff content and modified files are removed, primarily catching commits across repository mirrors.
The framework offers extensive filtering capabilities, allowing researchers to create custom datasets based on, among others, different programming languages, vulnerability classes, labeling strategies, source datasets, and temporal constraints. It also supports multiple data-splitting strategies, including random, temporal, and group-stratified splits that account for project identity by grouping related repositories to prevent undesired cross-split biases.
Using our framework, we systematically construct four datasets with increasing scope to evaluate the impact of data quality, quantity, and diversity on \spd\ performance. An overview of the size and label distribution for each dataset is shown in Table~\ref{tab:dataset-overview}. A detailed breakdown of all source datasets contributing to the respective target datasets can be seen in Table~\ref{tab:overview-datasets}, and the source set overlap of the five largest contributors in \ds{2} is shown in the alluvial diagram in Figure~\ref{fig:venn-ds2}.
\par\addvspace{\smallskipamount}\noindent
\textbf{\ds{1}\,--\,\cve-based \ccpp}: All \cve-mapped \ccpp\ commits, with additional benign commits to match the label distribution of \ds{2}.\par
\addvspace{\smallskipamount}\noindent
\textbf{\ds{2}\,--\,\ds{1} + Advisory-based \ccpp}: All advisory-mapped \ccpp\ commits, reflecting the sample distribution commonly used in \spd.\par
\addvspace{\smallskipamount}\noindent
\textbf{\ds{3}\,--\,\ds{2} + Automated tooling \ccpp}: All \ccpp\ commits, including those labeled using \ml\ approaches and traditional tools.\par
\addvspace{\smallskipamount}\noindent
\textbf{\ds{4}\,--\,All commits}: The largest dataset, removing all filtering constraints including the target programming language.\par
\addvspace{\smallskipamount}
To understand the structure and potential biases in \ds{2}, we analyze dataset provenance, project composition, token distributions, and latent representations. The project column (Table~\ref{tab:dataset-overview}) and treemap (Figure~\ref{fig:project-treemap}) reveal the imbalance in the distribution of commits to source repositories. Additionally, the hatched regions show that per-project \vfc\ ratios also vary considerably. This imbalance motivates split strategies that account for project identity and label ratios to prevent data leakage through project-specific patterns (Section~\ref{evaluation-experimental-results}) and balance each split.
\input{figures/framework/tsne-structure.tex}A t-SNE projection of CodeBERT~\citep{codebert} embeddings (Figure~\ref{fig:tsne-visualizations}) complements this picture. Samples cluster by source repository, consistent with project-specific coding and commit styles, and the clustering persists after fine-tuning on the detection task (Figure~\ref{fig:tsne-visualizations} (a) and (b)). When the projection is colored by commit timestamp, the temporal structure partially aligns with the project clusters, as expected since the source datasets crawled different repositories over different time periods. These visualizations are descriptive rather than causal, but they motivate evaluating models across project boundaries and questioning the reliability of temporal splits on aggregated corpora, both of which we test in Section~\ref{evaluation-experimental-results}.
The token-level analysis (Figure~\ref{fig:token-analysis}) shows that a substantial fraction of \diffs\ (\num{36}\% at \num{512} tokens, Table~\ref{tab:truncation-rates}) exceeds the sequence limit of smaller, commonly used encoder models, while the actual code changes account for only a modest fraction of each diff's total length. A significant portion of tokens consists of context lines, file headers, and syntax elements. This becomes particularly relevant when inputs must be truncated to fit model sequence limits, as naive end-truncation disproportionately discards vulnerability-relevant change tokens, unnecessarily impacting model performance~\citep{DBLP:conf/ndss/EvertzRNMNSGPSS26}. We therefore implement a context-aware truncation strategy that removes context lines in order of decreasing distance from the nearest code change, preserving the security-relevant signal while fitting within the model's token budget. As shown in Figure~\ref{fig:truncation-comparison}, this approach substantially reduces the fraction of discarded change tokens compared to naive truncation across all token limits. Beyond standard \diffs, we evaluate enriched input representations that replace the physically neighboring context lines with semantically targeted context derived from control-flow and data-flow analysis (Section~\ref{context-enrichment}).
\input{figures/dataset-characterization/token-distribution.tex}

%% file: figures/dataset-characterizations.tex
\begin{figure}[b]
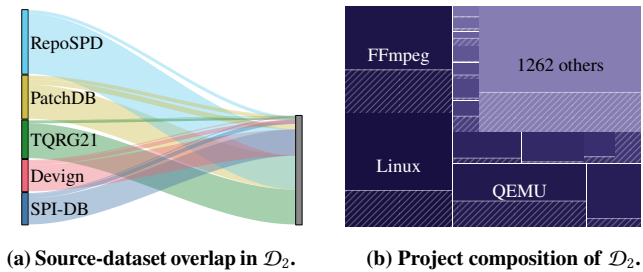

    \centering
    \begin{subfigure}[b]{0.45\linewidth}
      \centering
      \includesvgwithadjust{1.1}{-2pt 0pt 10pt 0pt}{data/dataset-characterization/dataset2-mr-advisory-cpp_source_overlap.svg}
      \caption{Source-dataset overlap in \ds{2}.}%
      \label{fig:venn-ds2}
    \end{subfigure}%
    \hfill
    \begin{subfigure}[b]{0.45\linewidth}
      \centering
      \includesvgwithadjust{1.05}{8pt 0pt 0pt 0pt}{data/dataset-characterization/dataset2-mr-advisory-cpp_project_treemap_vfc.svg}
      \caption{Project composition of \ds{2}.}
      \label{fig:project-treemap}
    \end{subfigure}
\caption{Dataset characterization for \ds{2}.
  (a)~Source-dataset overlap of the five largest contributors. (b)~Project treemap, hatched regions indicate its \vfc\ fraction.}%
\label{fig:dataset-characterization}%
\Description{Two panels. Left: alluvial diagram connecting the five largest source datasets to the deduplicated D2 dataset, with flows sized by the number of shared entries, showing substantial overlap between sources. Right: treemap of the projects in D2 where block area is proportional to the number of commits and hatched regions mark each project's share of vulnerability-fixing commits, showing that a few large repositories dominate the dataset.}%
\end{figure}%

%% file: figures/framework/tsne-structure.tex
\begin{figure}[b]
    \centering 
    \begin{subfigure}[b]{0.5\linewidth}
      \centering
      \includegraphics[width=0.7\linewidth]{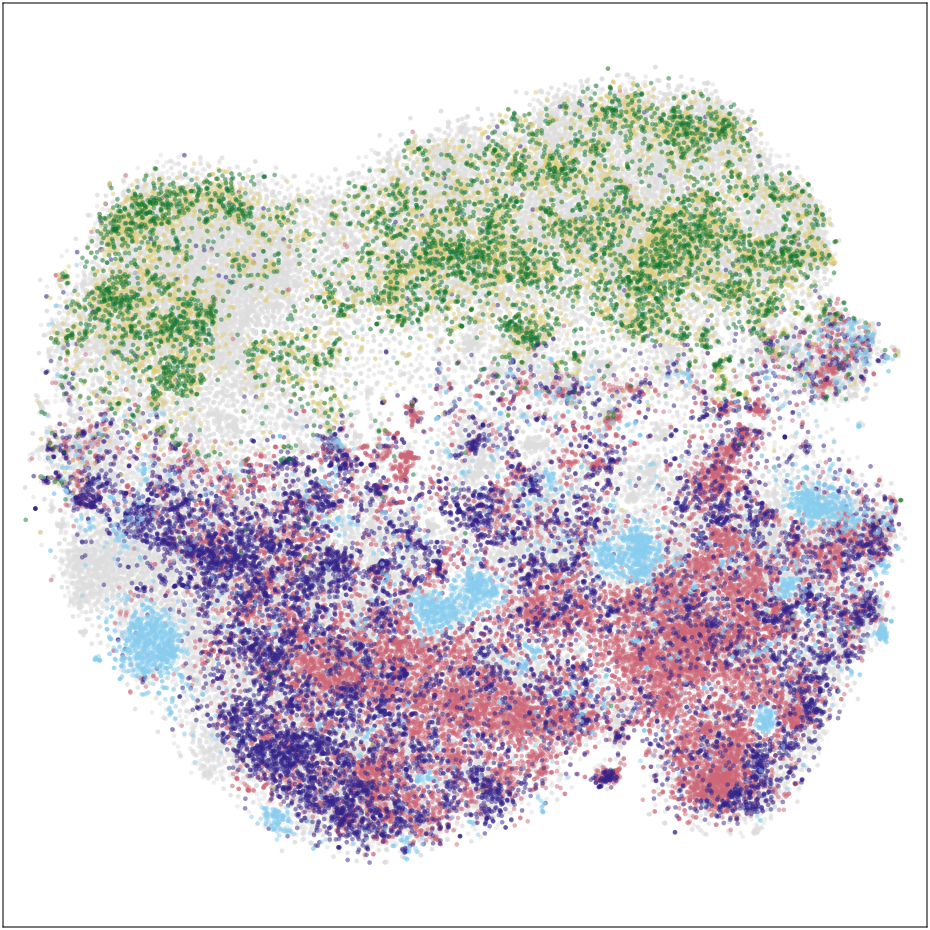}
      \caption{Stock, colored by project.}%
      \label{fig:tsne:stock-by-project}
    \end{subfigure}%
    \hfill
    \begin{subfigure}[b]{0.5\linewidth}
      \centering
      \includegraphics[width=0.7\linewidth]{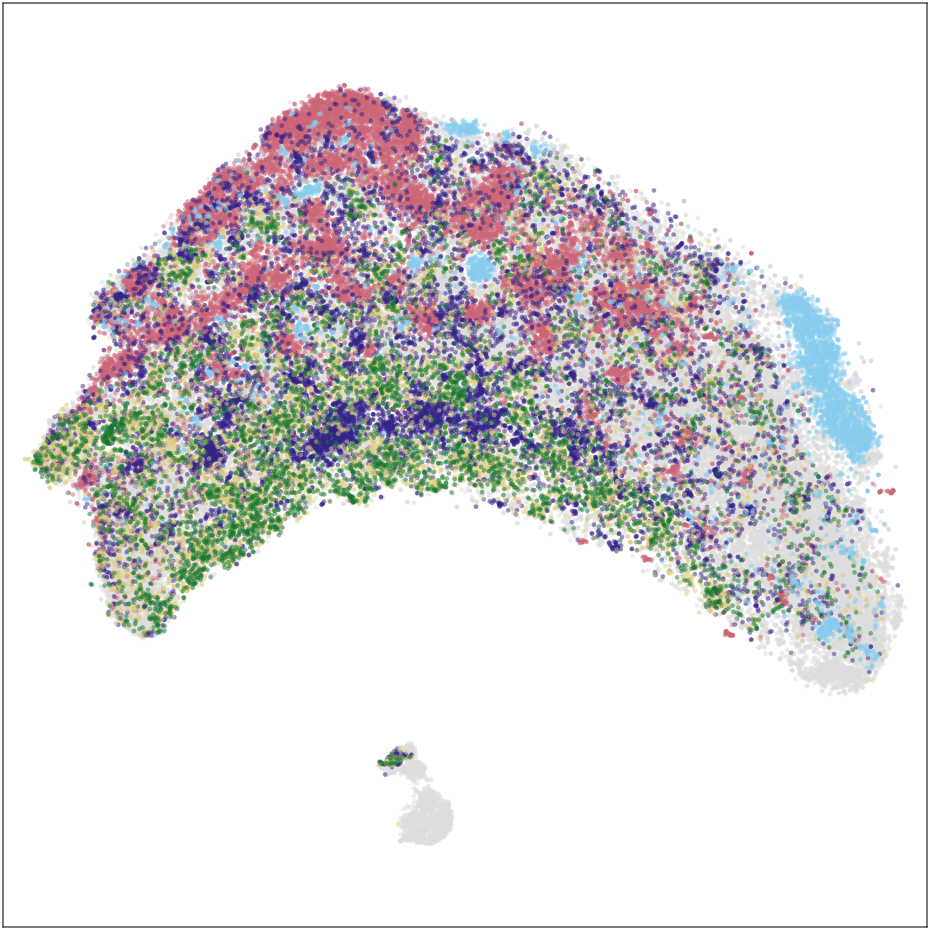}
      \caption{Finetuned, colored by project.}%
      \label{fig:tsne:finetuned-by-project}
    \end{subfigure}\\
    \begin{subfigure}[b]{0.5\linewidth}
      \centering
      \includegraphics[width=0.7\linewidth]{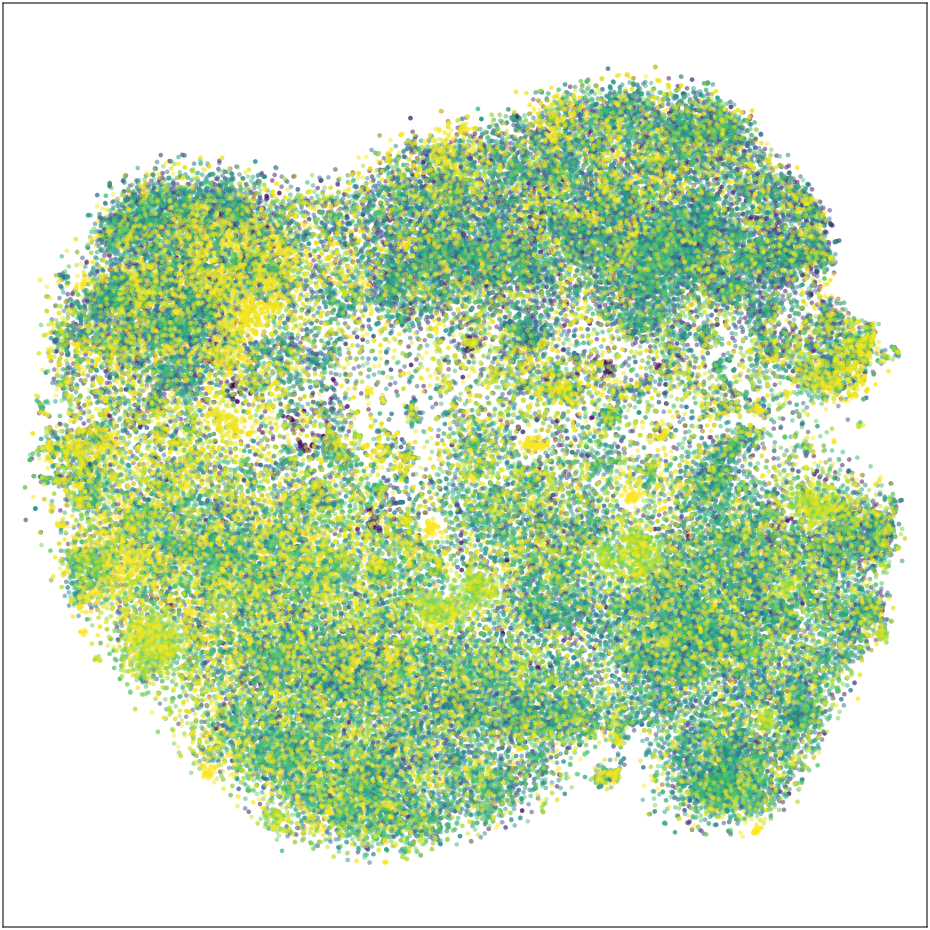}
      \caption{Stock, colored by timestamp.}%
      \label{fig:tsne:stock-by-timestamp}
    \end{subfigure}%
    \hfill
    \begin{subfigure}[b]{0.5\linewidth}
      \centering
      \includegraphics[width=0.7\linewidth]{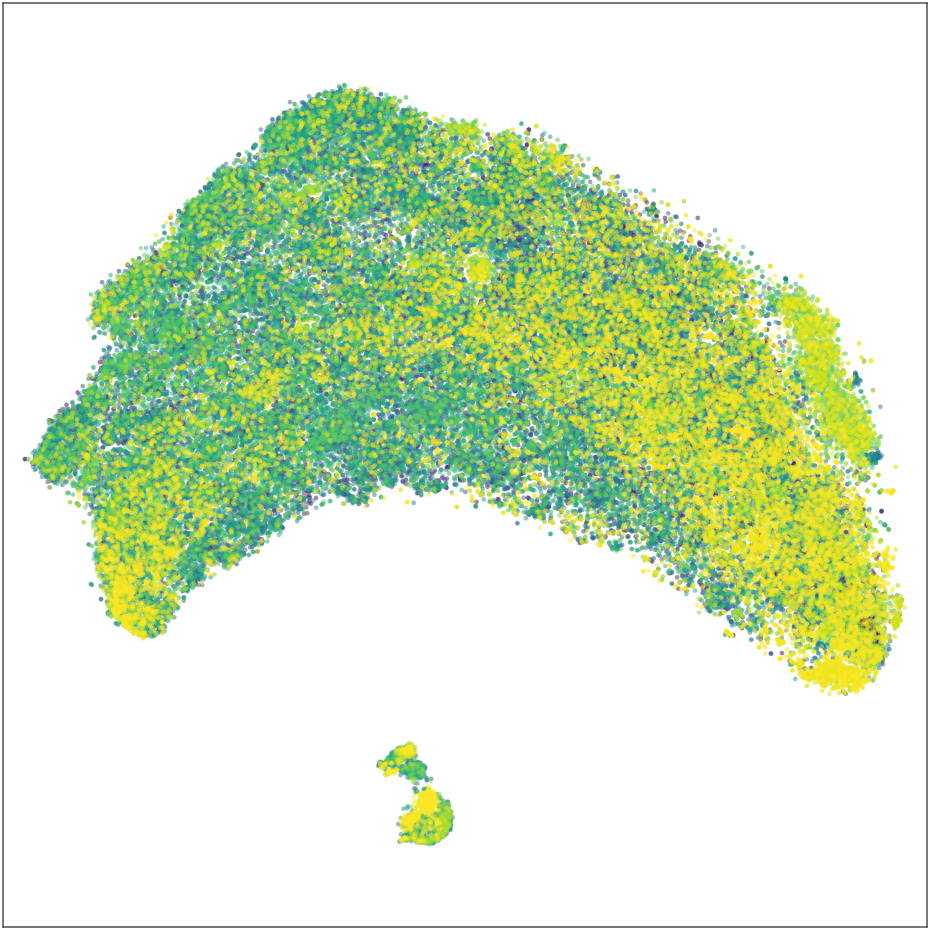}
      \caption{Finetuned, colored by timestamp.}%
     \label{fig:tsne:finetuned-by-timestamp}
    \end{subfigure}%
    \caption{t-SNE visualizations of \ds{2} based on CodeBERT~\cite{codebert} embeddings.In the top row, points are colored by the five most frequent source repositories, remaining projects in gray. In the bottom row, lighter colors represent more recent commits.
  The model is fine-tuned on the \diff.}%
\label{fig:tsne-visualizations}
\Description{Four t-SNE scatter plots of CodeBERT embeddings of the D2 dataset, arranged in two rows and two columns. The left column uses stock CodeBERT, the right column a fine-tuned model. The top row colors points by the five most frequent source repositories and shows pronounced repository clusters in both columns. The bottom row colors points by commit timestamp without a comparable cluster structure.}
\end{figure}

%% file: figures/dataset-characterization/token-distribution.tex
\begin{figure}[t]
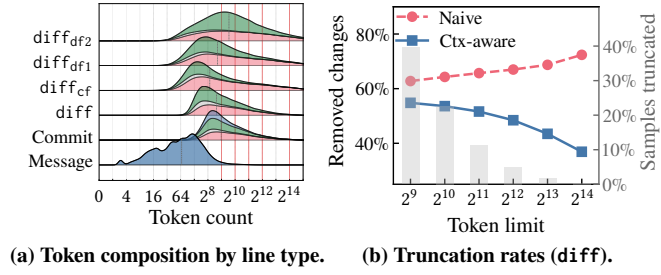

      \centering
          \begin{subfigure}[b]{0.49\linewidth}
      \centering
       \includesvgwithadjust{1.1}{15pt 0pt 0pt 0pt}{data/dataset-characterization/dataset2-mr-advisory-cpp_token_distribution_stacked.svg}
    \caption{Token composition by line type.}%
      \label{fig:token-distribution}
    \end{subfigure}
        \begin{subfigure}[b]{0.49\linewidth}
      \centering
       \includesvgwithadjust{1.1}{10pt 7pt 10pt 5pt}{data/dataset-characterization/dataset2-mr-advisory-cpp_truncation_comparison.svg}
      \caption{Truncation rates (\diff).}%
      \label{fig:truncation-comparison}
    \end{subfigure}
    \caption{Token-level analysis of \ds{2}.
      (a)~Token distributions decomposed into changed lines (red), file headers (gray), context (green), and commit message (blue). (b)~Fraction of change tokens among discarded tokens under naive vs.\ context-aware truncation. Gray bars show affected samples (right axis).}%
    \label{fig:token-analysis}
    \Description{Two subfigures. Left: six stacked ridge plots showing
token distributions for different diff representations, decomposed by
line type (changes, syntax, context, message), with vertical reference
lines at common model sequence limits. Right: line chart comparing
naive and context-aware truncation showing
that context-aware truncation consistently removes a smaller proportion
of change tokens (51--56 percent versus 70--82 percent), with gray
bars indicating the decreasing number of affected samples at higher
limits.}
\end{figure}

%% file: sections/Evaluation.tex
Our evaluation asks whether fine-tuned code LMs learn transferable, security-relevant representations of code changes, whether model capacity or additional code context can strengthen that code signal, or whether reported \spd\ performance is shaped primarily by how these systems are trained and evaluated. We decompose this question into three research questions and study each in turn.
\par\addvspace{\smallskipamount}\noindent
\rqlabel{one}{RQ1\,--\,Training \& Evaluation Protocol.} How do splitting strategies and additional natural language context influence \spd\ performance and generalizability?\par
\addvspace{\smallskipamount}\noindent
\rqlabel{two}{RQ2\,--\,Model \& Context Impact.} How do model scale and additional code context impact \spd\ performance?\par
\addvspace{\smallskipamount}\noindent
\rqlabel{three}{RQ3\,--\,Label \& Dataset Quality.} How accurate are the labels, and how do label quality and dataset diversity impact model training and evaluation?\par
\addvspace{\smallskipamount}
\subsection{Models and Training Setup}
We evaluate popular code LMs (Table~\ref{tab:model-overview}) using \num{60}/\num{20}/\num{20} training, validation, and test splits. The comparatively large evaluation partitions ensure sufficient test samples under group-stratified conditions, where repository constraints reduce the effective sample size. We train all fully fine-tuned models for \num{10} epochs and LoRA fine-tuned models for \num{5} epochs and select the best validation F1 checkpoint for testing. For group-stratified splits, related repositories (e.g., forks) are grouped so that all commits from related projects always appear in the same partition, preventing cross-split data leakage. Groups are assigned to splits via a greedy algorithm refined by local search to approximate the target ratios. The assignment proceeds hierarchically by first separating test (\num{20}\%) from train and validation (\num{80}\%) and then splitting the remainder into train (\num{60}\%) and validation (\num{20}\%). All experiments are run on NVIDIA A100, H100, and H200 GPUs.
All models are trained using an effective batch size of \num{64}, AdamW~\citep{DBLP:conf/iclr/LoshchilovH19}, and a learning rate of \num{2e-5}, with the classification heads provided by the Hugging Face \texttt{*ForSequenceClassification} model implementations. To train Qwen2.5-Coder~\citep{Qwen25:DBLP:journals/corr/abs-2412-15115} and Qwen3-Coder-Next~\citep{DBLP:journals/corr/abs-2603-00729}, we use low-rank adaptation~\citep{Lora:DBLP:conf/iclr/HuSWALWWC22} with rank \num{16} and $\alpha=\num{32}$. Class imbalance is handled via inverse frequency weighting. 
We report F1 scores as a well-recognized metric and adapt the vulnerability detection score (VD-S) introduced by~\citet{PrimeVul} as the patch detection score PD-S ($=$ VD-S) $= FNR @ (FPR \leq r)$ where $r\in[0,1]$ to quantify detection rates under constrained misclassification rates to help mitigate the base rate fallacy~\citep{DBLP:conf/uss/ArpQPWPWCR22}. Following the evaluation protocol of~\citet{PrimeVul}, we set $r=\num{0.005}$. For random and group-stratified splits, we report mean and standard deviation across three seeds unless stated otherwise. The temporal split is inherently unique.
\input{tables/model-overview}\subsection{Input Representations and Context Enrichment}\label{context-enrichment}
\input{figures/framework/sample-commit.tex}Standard \diffs\ from version control systems capture what was modified but may lack sufficient semantic context for accurate classification. Consider the memory-leak fix in Figure~\ref{fig:ffmpeg-diff}, where the added and deleted lines specify what was modified, but without additional context, the impact remains unclear. To clean the \diffs, we strip comments by retrieving the full file versions before and after each commit and using Tree-sitter~\citep{tree-sitter} to parse them into concrete syntax trees, precisely identifying comment nodes rather than relying on regex heuristics that cannot distinguish comments from comment-like patterns in string literals.
To systematically evaluate whether additional semantic context improves \spd, we implement a lightweight intra-procedural context enrichment method. For each changed function, we generate concrete syntax trees for both pre- and post-patch versions using Tree-sitter and compute a structural diff using GumTree~\citep{GumTree}. From the changed nodes, we identify relevant statements through forward and backward slicing along definition-use chains at depths $d\in\{1,2\}$, and augment the result with immediate control-flow enclosures (if-statements, loops). This produces three progressively enriched representations: control-flow context (\texttt{cf}), single data-flow pass (\texttt{df}\textsubscript{1}), and two data-flow passes (\texttt{df}\textsubscript{2}). Each context level is colored accordingly in Figure~\ref{fig:ffmpeg-diff}. Algorithm~\ref{alg:context-enrichment} in Appendix~\ref{app:context-alg} formalizes the procedure.
Our method averages $\num{1.72}$ seconds per sample on \num{500} PatchDB~\citep{PatchDB} samples, compared to $\num{57.36}$ seconds for the Joern-based~\citep{Joern} analysis used by GraphSPD~\citep{GraphSPD}. Unlike standard unified diffs that include physically neighboring context lines regardless of their semantic relevance, the enriched representations replace these with targeted context based on control-flow and data-flow dependencies. As a result, the control-flow representation (\texttt{cf}) has a median size of \num{280} tokens\footnote{Token counts use the CodeBERT tokenizer but are similar across tokenizers.}, smaller than the baseline \diff\ at \num{394} tokens ($0.71\times$), because non-dependent neighboring lines are removed and only the enclosing control-flow structures are retained. Adding data-flow context through two incremental data-flow passes increases sizes to \num{417} (\texttt{df}\textsubscript{1}, $1.06\times$) and \num{747} (\texttt{df}\textsubscript{2}, $1.9\times$) tokens. Table~\ref{tab:truncation-rates} reports the resulting truncation rates across all representations and model sequence limits. The data-flow representations face higher truncation pressure at small limits, and all enriched representations retain a heavier tail at larger ones.
\input{tables/truncation-rates}
\subsection{Experimental Results}\label{evaluation-experimental-results}%
We structure our evaluation along our three research questions. 
We first quantify the relative contributions of commit messages and code changes and evaluate how splitting strategies affect performance estimates (\rqref{one}{RQ1}). 
These protocol effects are demonstrated on \ds{2}, whose composition reflects the sample distribution commonly used in \spd\ research, ensuring that the resulting protocol recommendations directly apply to current practice. 
Using group-stratified splits, we then assess the impact of model capacity and context enrichment across seven models on both \ds{2} and its \cve-mapped counterpart \ds{1} (\rqref{two}{RQ2}). 
Finally, we quantify the label accuracy of both datasets and examine its impact on evaluation and training through a manual error analysis and cross-dataset training, complemented by an exploratory comparison to prompt-based classification (\rqref{three}{RQ3}).
To contextualize model performance, we follow existing studies~\cite{TopScoreWrongExam:Risse2025,VulnPatchPairs} and include a baseline that reveals how much performance is attainable from surface lexical statistics alone. 
Specifically, we train a logistic regression on term frequency-inverse document frequency (TF-IDF) features of each respective input signal.
\input{tables/rq1-2-performance}\rqsection{one}{Training and Evaluation Protocol}\setrq{one}%
Following existing observations on natural language (\texttt{NL}) reliance~\citep{CompVPD:chen2024compvpditerativelyidentifyingvulnerability, DBLP:conf/icse/SunXLXZHZ23}, we quantify the relative contributions of the commit message (\texttt{NL}), code comments (\texttt{NL}), and the \diff\ itself (\texttt{PL}) for \spd.
We follow common evaluation protocols and train \codebert~\citep{codebert} and \ctfive~\citep{CodeT52021} on \ds{2} under random split conditions (Table~\ref{tab:rq1-2-performance}).
Models are always trained and evaluated on the same representation, rather than withholding inputs at test time.
Removing the commit message drops F1 by roughly a quarter while PD-S more than doubles, indicating a substantial loss in detection capability.
Conversely, training on the commit message alone yields performance within a few percentage points of the full-commit baseline, and both patterns persist directionally across all splitting strategies.
This indicates that models primarily exploit the textual signal for classification, masking any code reasoning capability, consistent with findings by \citet{CompVPD:chen2024compvpditerativelyidentifyingvulnerability} and \citet{DBLP:conf/icse/SunXLXZHZ23}.
Removing comments within \diffs\ only marginally reduces performance on this dataset.
Training dynamics for all experiments (Figure~\ref{fig:training-dynamics}) underline the training stability across representations.
Though training sets differ, full-commit performance is in line with or above comparable message-based systems~\citep{SPI,10.1145/3749370,DBLP:journals/corr/abs-2308-15233,CompVPD:chen2024compvpditerativelyidentifyingvulnerability}.
The lexical baseline follows the same pattern, although its PD-S starts poorly and rises less sharply, strengthening existing evidence of reliance on the commit message and, to a lesser extent, comments.
Relying on the message is also methodologically circular, as several of the underlying source datasets identify \vfcs\ through keywords or \cve\ identifiers in commit messages, so message-based classification partly reproduces the labeling heuristic rather than detecting the fix.
In deployment, the same reliance breaks down for silent security patches, whose messages typically omit exactly these cues.
Since our goal is to understand the code reasoning capabilities of these models, all subsequent experiments use only the code changes and their contextual signals.
\par
\input{figures/evaluation/temporal-sensitivity.tex}
\input{figures/evaluation/model-context-forest.tex}Beyond the reliance on message patterns, evaluation must account for data leakage and spurious correlations, which can cause severe overestimation of performance~\citep{DBLP:conf/uss/ArpQPWPWCR22}, and for distribution shift over time~\citep{BeyondTesseract:Chow2026,DBLP:journals/corr/abs-2402-01359}. A t-SNE analysis of \codebert\ embeddings reveals clear repository-specific clusters in \ds{2} (Figure~\ref{fig:tsne-visualizations}), which together with reported cross-project performance drops for \vulndet~\citep{DBLP:conf/icse/SteenhoekRJL23} motivates an evaluation across project boundaries.
Compared to random splits, group-stratified splits, which keep all commits of related repositories within a single partition, reduce F1 on the code-only representations by \num{16}-\num{28}\% across all models, including the lexical baseline (Table~\ref{tab:rq1-2-performance}). For \diff-only inputs, PD-S saturates at \num{0.97}--\num{0.99}, leaving essentially no detection on unseen projects under the strict false positive budget. Both trends confirm that classifiers pick up project-specific patterns, including purely lexical ones.
Temporal splits degrade further, with commit-level F1 dropping from \num{0.88} to \num{0.58} for \codebert.
To understand this temporal sensitivity, we conduct a sliding window analysis following \citet{BeyondTesseract:Chow2026}, training a fresh \codebert\ model at each window position (Figure~\ref{fig:temporal-sensitivity}). As the window advances, the training and test project distributions drift apart, the share of test commits from unseen projects grows, and the test vulnerability rate declines as later periods contain fewer advisory-linked commits. Test F1 tracks these trends and falls from \num{0.62} to \num{0.29} although only the date boundary is shifted. While part of this decline follows the declining vulnerability rate, the persistently high PD-S indicates that temporal performance on these aggregated datasets is dominated by dataset composition rather than genuine temporal shift.
Robustness to temporal shift remains an important property for deployed detection systems, but evaluating it faithfully requires a temporally stable dataset composition that aggregated \vfc\ corpora do not currently provide.
We therefore adopt group-stratified splitting for all subsequent experiments.
\begin{finding}
\findhead{RQ1} Confirming prior observations, classifiers rely on natural language signals when present. Understanding code reasoning capabilities therefore requires code-only evaluation, for which only group-stratified splits provide faithful estimates on unseen projects.
\end{finding}
\rqsection{two}{Model and Context Impact}\setrq{two}%
To investigate the impact of model architecture, capacity, and enriched context on \spd\ performance, we evaluate all models from Table~\ref{tab:model-overview} across the context levels of Section~\ref{context-enrichment} on \ds{2} and \ds{1}. \ds{2} serves as the scaled dataset whose composition matches commonly reported evaluations, while \ds{1} restricts positives to \cve-mapped commits at the cost of training volume. The impact of this difference is examined in \rqref{three}{RQ3}. Mean F1 and PD-S with standard deviations over three group-stratified split seeds are shown in Figure~\ref{fig:rq2:model-vs-context}.
The encoder-based models exceed the lexical baseline only marginally on both datasets, and neither specialized pretraining (CodeBERT C++, CommitBART) nor doubled context length (UniXcoder) provides measurable gains. Only the larger decoder-based Qwen models show clear improvements, especially on \ds{1}. To separate model capacity from the Qwen models' longer inputs, we retrain Qwen3-Coder-Next with \diff\ inputs truncated to \num{512} instead of \num{16384} tokens and observe mean differences below a third of the standard deviation. This suggests that capacity rather than input length drives the improvement, and raises the question of why untruncated inputs do not help. To test whether locality-based context underuses the available tokens, we next replace it with semantically selected statements.
\par
\input{tables/repospd-baselines}
We evaluate the enriched representations of Section~\ref{context-enrichment} alongside RepoSPD~\cite{RepoSPD}, a recent system that incorporates inter-procedural and repository-level context (Table~\ref{tab:repospd-baselines}). As RepoSPD's graph construction does not succeed for every commit, all systems are compared on the \num{77}--\num{85}\% of samples per split that survive its preprocessing, marked with an asterisk. Surprisingly, no intra-procedural context level yields a consistent improvement for any model on either dataset. Inter-procedural repository context yields moderate F1 gains over the UniXcoder backbone, but these do not survive the strict false positive budget. RepoSPD reproduces its originally reported results on the PatchDB subset. On \ds{1}*, RepoSPD raises F1 by \num{0.07} over its UniXcoder backbone, yet its PD-S remains at the level of the backbone on the plain \diff\ and of the lexical baseline, despite preprocessing costs of over \num{5200} CPU hours on \ds{2}. Notably, the fine-tuned Qwen3-Coder-Next surpasses RepoSPD in F1 on every subset and clearly in PD-S on \ds{1}*, operating on the plain \diff\ without any repository context. While the added statements provide semantically relevant context, models appear unable to exploit them. An integrated gradients analysis~\citep{IG:DBLP:conf/icml/SundararajanTY17} (Figure~\ref{fig:attention-attribution}) is consistent with this, as the changed lines receive no larger attribution share than surrounding context or file headers and the added context does not shift this distribution. Since attribution reflects input sensitivity rather than causal reliance, we treat it as a supporting diagnostic. We discuss the relevance of context for \spd\ in Section~\ref{sec:threats-to-validity}.
\begin{finding}
\findhead{RQ2} Larger decoder-based models improve detection, but even the largest model misses most fixes under a strict false positive budget. Neither the evaluated intra-procedural enrichments nor inter-procedural repository context yields gains that survive this budget.
\end{finding}
\rqsection{three}{Label and Dataset Quality}\setrq{three}%
Label quality is a known issue for \spd\ datasets and has been quantified especially for \vulndet~\cite{PrimeVul,DiverseVul}. However, \vulndet\ commonly extracts single pre-fix functions labeled according to the \spd\ label, and the resulting error rates do not transfer to \spd\ due to a mismatch in scope. A commit fixing a memory bug that depends on a function's input is a clean function-level \vulndet\ target, yet if that input is not attacker-controlled, the commit constitutes hardening rather than a \vfc.\footnote{E.g., \url{https://github.com/varnishcache/varnish-cache/commit/2f10ef7d8a1b} widens an undersized \texttt{memset} that can only under-initialize memory, which is in turn fully re-initialized before every use. The commit nevertheless carries a \vfc\ label.} Conversely, a commit introducing guards along the data flow to a security-critical sink is a valid security patch, while the extracted pre-fix function may not contain the sink and appears mislabeled for \vulndet.\footnote{E.g., \url{https://github.com/heimdal/heimdal/commit/f9ec7002cdd5} resolves CVE-2021-44758 by fixing a guard in the changed function, while the NULL dereference it prevents executes in an unchanged function.}
\par
\input{figures/evaluation/attention-attribution.tex}To quantify the label accuracy of \ds{1} and \ds{2}, we randomly sample \num{200} commits from \ds{2}, split equally among benign commits, \vfcs\ with an associated \cve, and \vfcs\ without one. Two security experts independently assessed each commit and a third expert adjudicated the \num{11} disagreements (raw agreement \num{94.4}\%, Cohen's $\kappa = 0.89$~\citep{CohenKappa}). Table~\ref{tab:manual-label-analysis} summarizes the results. Benign and \cve-associated commits show label accuracies above \num{90}\%, yielding a population-weighted accuracy of \num{92}\% for \ds{1}. \vfcs\ without \cve\ association, however, reach a correctness of only \num{28.8}\% when requiring both a security-relevant pattern and reasonable in-context exploitability, and \num{66.7}\% when the pattern alone suffices. Overall, the population-weighted label accuracy of \ds{2} is estimated at \num{79.8}\% under the strict definition. We return to the implications of these two notions of a security fix in Section~\ref{sec:threats-to-validity}.
\par
\input{tables/manual-label-analysis}
To understand how this noise affects reported performance and where models actually fail, we manually analyze \num{200} misclassifications of the largest model, Qwen3-Coder-Next, on the standard \diff. From each dataset we draw \num{50} random false positives (FP) and \num{50} random false negatives (FN). Each case is attributed to either a label or a model failure, further divided by whether a security-relevant pattern is present in the \diff\ (Figure~\ref{fig:error-analysis}). For false negatives, the two datasets differ sharply. On \ds{2}, the majority of misses are mislabeled positives, so model predictions are heavily penalized by the labeling issues quantified above. Conversely, higher scores on \ds{2} can only be achieved by reproducing these mislabels, so the strong results of message-based models in \rqref{one}{RQ1} at least partially reflect the labeling process rather than security understanding. On \ds{1}, most false negatives are model-attributable cases without a common security pattern in the \diff, making missing context the key factor. For false positives, both datasets show a similar distribution as they share their benign commits. Notably, the rate of mislabeled samples among them exceeds the \num{7.6}\% observed on randomly drawn benign commits, most clearly on \ds{1}, suggesting that models have to some extent learned to identify security-relevant commits. The remaining model failures largely exhibit security-relevant patterns that are deemed non-exploitable, often fixes for fuzzing findings. 
\begin{finding}
\findhead{RQ3} Label error concentrates in positives lacking \cve\ association and primarily distorts evaluation, where models are penalized for correctly rejecting mislabeled samples. On \cve-associated labels, models miss fixes whose evidence lies outside the \diff\ and flag benign changes that carry familiar security cues.
\end{finding}
\input{figures/evaluation/error-analysis.tex}
\input{figures/evaluation/training-dynamics.tex}
\par
Having isolated the evaluation impact of label noise, we next examine its impact during training by training Qwen3-Coder-Next on all four dataset compositions while validating and evaluating on a single group-stratified split of \ds{1}. Each training pool is filtered to exclude all repository groups present in the \ds{1} validation and test partitions. While F1 remains largely stable across compositions (Table~\ref{tab:cross-dataset-cve}), PD-S deteriorates as the non-\cve\ positives of \ds{2} and \ds{3} are added, and only recovers to the level of training on \ds{1} alone with \ds{4}, which contributes additional \cve-mapped commits across all languages. Under strict false positive budgets, \cve-mapped training samples thus outweigh sheer volume in this experiment.
\par\textbf{Prompt-based Classification\;}
As an exploratory outlook, we lastly consider prompt-based classification, which forms the foundation of the agentic systems increasingly adopted across software engineering~\citep{Agent4SE:LiuTOSEM2026}, a direction we revisit in Section~\ref{sec:threats-to-validity}. We evaluate GLM-4.7~\citep{GLM45:DBLP:journals/corr/abs-2508-06471} (\num{355}B parameters) with activation-aware weight quantization (AWQ)~\citep{DBLP:conf/mlsys/0002TTYCWXDG024} on \ds{1} without any fine-tuning. Since chain-of-thought prompting is not universally beneficial~\citep{DBLP:conf/iclr/SpragueYRJWSZYM25}, we compare direct answer (DA) classification and chain-of-thought (CoT) reasoning across all context levels (Table~\ref{tab:prompt-based}). We report only F1, as the discrete outputs do not allow calibrating the false positive rate required for PD-S. The results further constitute an upper bound on the intrinsic capability, as we cannot rule out that samples or their accompanying advisories were part of the pretraining data (Section~\ref{sec:threats-to-validity}). An empirical contamination probe restricted to \cves\ published after the model's training cutoff is not feasible, as \ds{1} contains only \num{18} newer \vfcs. DA outperforms CoT, and both perform on par with the smaller fine-tuned models on \ds{1}, while the fine-tuned Qwen models achieve higher performance despite their smaller size.
\input{tables/cross-dataset-cve.tex}
\subsection{Discussion and Threats to Validity}%
\label{sec:threats-to-validity}%
\par\textbf{Discussion\;}
\par\textit{a) \spd\ scope.\;}%
\spd\ is commonly framed as a binary or ranking task where a model classifies a given commit as security relevant or identifies the relevant commit within a candidate set. However, our label analysis (Table~\ref{tab:manual-label-analysis}) reveals that current labels inconsistently mix generally security-relevant changes with fixes for exploitable vulnerabilities. Even on \ds{1}, where labels align more closely with exploitability, the error analysis (Figure~\ref{fig:error-analysis}) shows that models produce false positives when a familiar security cue is present and false negatives when it is absent. Both notions of a security fix are worth detecting, and which one a system should target depends on how its flags are consumed. Automated dependency updates can act on the broader notion at little cost, whereas alerts that trigger expert triage or costly patching processes require confidence in exploitability. We therefore argue that authors should state which definition their labels capture and which their system targets, and choose datasets, inputs, and evaluation criteria accordingly. 
\par\textit{b) Context relevance.\;}%
Our experiments show that additional code context does not resolve the limitations of code-only \spd. Intra-procedural enrichment yields no consistent benefit, and on \ds{1}* the PD-S of RepoSPD remains at the level of its UniXcoder backbone on the plain \diff\ and even of the lexical baseline (Table~\ref{tab:repospd-baselines}). At the same time, the error analysis attributes \num{54}\% of false negatives on \ds{1} to the absence of a decisive signal in the \diff\ alone. This suggests that the limiting factor is not the amount of context but whether the decisive context is selected and effectively used. Judging whether a change prevents exploitation requires knowledge of the underlying system, its interactions, and the capabilities of an attacker, and during our manual analyses the majority of time was spent recovering exactly this context. Group-stratified evaluation mirrors deployment on unseen projects, where such knowledge must be acquired rather than memorized. Systems could achieve this through retrieval that grounds each change in its repository, through codebase-specialized models that track a project and classify its commits as they arrive, or through agentic systems that gather the required context on demand. Evaluation of such approaches on public benchmarks, however, is subject to the contamination risks discussed in \textit{v)}.  
\input{tables/prompt-based}\par\textbf{Threats to Validity\;}
\par\textit{i) Dataset bias.\;}%
Dataset curation directly shapes which patterns a model can learn. Our main evaluations are limited to \ccpp\ and therefore to the security issues common in these languages. Some transfer to languages with similar issues appears plausible, and adding \cve-mapped commits from other languages slightly improved results on \ccpp\ (Table~\ref{tab:cross-dataset-cve}), but we have not evaluated models on other languages. In addition, \ds{1} over-represents publicized vulnerabilities in large, well-audited codebases. Systems trained and evaluated on it therefore should be careful to claim the ability to detect silent security fixes without additional evaluation, as such fixes likely follow a different distribution. Still, a model that performs well on \ds{1} would be valuable in practice. It could monitor repositories and flag vulnerability-fixing commits before the corresponding \cve\ is announced, shortening the median \num{25}-day exposure window between patch release and advisory publication~\citep{DBLP:journals/tse/ImtiazKW23}. 
\par\textit{ii) Split bias.\;}%
Reported performance depends directly on the splitting strategy. Group-stratified splits estimate performance on unseen projects, matching the scenario of monitoring novel repositories. Within-project splits remain interesting for systems that continuously observe a known project, complementing the context discussion in \textit{b)}. However, due to the composition of the existing datasets (Figure~\ref{fig:dataset-characterization}~(b)), creating well-balanced splits is not trivial, as individual repositories make up large parts of each split. We mitigate this by enforcing class ratios during group assignment and by reporting mean and standard deviation over three split seeds. 
\par\textit{iii) Residual label error.\;}%
Even on \ds{1}, residual label errors remain and appear as label-attributable misclassifications in the error analysis (Figure~\ref{fig:error-analysis}). Benign-side noise is particularly relevant for PD-S, as mislabeled benign samples that a model correctly flags consume the strict FPR budget and tighten the decision threshold. PD-S therefore represents a conservative estimate of the actual detection capability. Ideally, benign samples should be curated to reflect realistic commit distributions while ensuring that they contain no silent patches. 
\par\textit{iv) Model selection and parameter tuning.\;}%
We selected models that reflect commonly used architectures in \spd\ and \vulndet\ across several generations and sizes, yet the results cover only a small sample of existing models. We follow common design principles for the classification heads and the LoRA setup with widely used training hyperparameters, but perform no per-model hyperparameter tuning, which would likely yield marginal improvements. Considering the stable convergence observed during training (Figure~\ref{fig:training-dynamics}), we believe the observed trends support comparative evaluation.
\par\textit{v) Training leakage for pretrained models.\;}%
Pretrained models inherit information from their pretraining corpora and potential leakage must be considered. For \spd\ we distinguish three channels: (1) source code of the evaluated repositories providing repository-level knowledge, (2) commits directly leaking test samples, and (3) vulnerability information from advisories, bug trackers, or security mailing lists providing label information. \codebert\ and \ctfive\ are pretrained on CodeSearchNet~\citep{DBLP:journals/corr/abs-1909-09436}, which contains no \ccpp\ code, and are therefore unaffected. CodeBERT C++ has seen some of the evaluated repositories (1), UniXcoder includes a crawled web corpus that likely contains vulnerability information (3) and CommitBART is trained directly on commits, though without explicit \spd\ labels (2). For the large generative models (Qwen2.5-Coder, Qwen3-Coder-Next, GLM-4.7), contamination across all three channels is not verifiable but very likely. We also explored prompt-based experiments with frontier models but repeatedly observed that Claude Opus 4.7 responds to \diff-only inputs, without the use of external tools such as web browsing, with the correct \cve\ or commit ID. We therefore consider all results of large generative models an upper bound of their intrinsic performance. 

%% file: tables/model-overview.tex
\begin{table}[th]
\small
\caption{Overview of evaluated models.}
\centering
\setlength{\tabcolsep}{5pt}
\begin{threeparttable}
\begin{tabular}{l | r | r | l}
\toprule
\textbf{Model} & \textbf{Param} & \textbf{Tokens} & \textbf{Arch} \\
\midrule
CodeBERT~\citep{codebert} & \num{125}M & \num{512} & Enc \\
UniXcoder~\citep{unixcoder} & \num{125}M & \num{512}/\num{1024} & Enc-Dec \\
CodeBERT C++~\citep{codebert-cpp} & \num{125}M & \num{512} & Enc \\
CodeT5-Large~\citep{CodeT52021} & \num{770}M & \num{512} & Enc-Dec \\
CommitBART~\citep{commit-bart} & \num{140}M & \num{1024} & Enc-Dec \\
Qwen2.5-Coder~\citep{Qwen25:DBLP:journals/corr/abs-2412-15115} & \num{14}B & \num{4096}\tnote{1} & Dec \\
Qwen3-Coder-Next~\citep{DBLP:journals/corr/abs-2603-00729} & \num{80}B-A\num{3}B & \num{16384}\tnote{1} & Dec \\
\midrule
GLM-4.7~\citep{GLM45:DBLP:journals/corr/abs-2508-06471}\tnote{2} & \num{355}B & \num{8192}\tnote{1} & Dec \\
\bottomrule
\end{tabular}
\begin{tablenotes}
\small
\item[1] The model's max capacity is larger.
\item[2] Evaluated in a prompt-based setting.
\end{tablenotes}
\end{threeparttable}
\vspace{-3mm}%
\label{tab:model-overview}
\end{table}

%% file: figures/framework/sample-commit.tex
\begin{figure}[th]
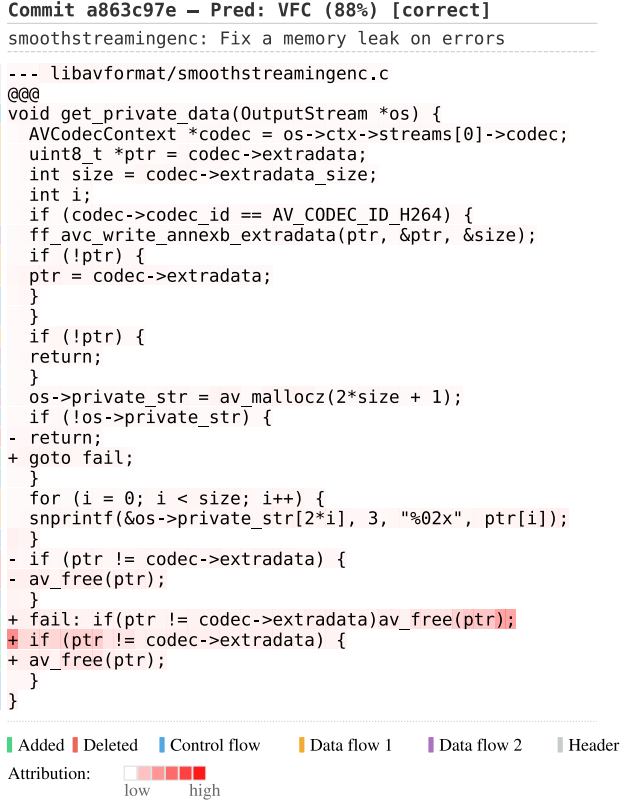

  \centering
  \includesvgwithadjust{1}{0pt 0pt 0pt 0pt}{data/attribution_diff_a863c97e.svg}
  \caption{Example \vfc\ from FFmpeg with integrated gradients attribution (absolute values) overlaid on the enriched \diff.}%
  \label{fig:ffmpeg-diff}
     \Description{An example vulnerability-fixing commit from FFmpeg showing a code diff with integrated gradients attribution overlaid. Edge markers distinguish added and deleted lines from control-flow and data-flow context levels, as labeled in the embedded legend. Red shading indicates attribution magnitude per token.}
\end{figure}%

%% file: tables/truncation-rates.tex
\begin{table}[t]
\caption{Share (\%) of \ds{2} samples exceeding common model sequence limits per input representation. Values for \num{512} and \num{1024} tokens are averaged over the tokenizers of the four and two models sharing each limit, with standard deviation.}
\centering
\setlength{\tabcolsep}{4pt}
\begin{tabular}{l | r r r r r}
\toprule
\textbf{Repr.} & \textbf{512} & \textbf{1024} & \textbf{4096} & \textbf{8192} & \textbf{16384} \\
\midrule
\diff & $35.8 \pm 4.4$ & $15.7 \pm 1.4$ & \num{2.4} & \num{0.7} & \num{0.1} \\
\texttt{cf} & $35.1 \pm 1.2$ & $23.2 \pm 0.3$ & \num{9.4} & \num{5.3} & \num{2.7} \\
\texttt{df}\textsubscript{1} & $43.5 \pm 1.5$ & $27.5 \pm 0.8$ & \num{10.5} & \num{5.9} & \num{3.0} \\
\texttt{df}\textsubscript{2} & $58.6 \pm 1.5$ & $38.4 \pm 1.4$ & \num{13.7} & \num{7.5} & \num{3.7} \\
\bottomrule
\end{tabular}
\vspace{-3mm}%
\label{tab:truncation-rates}
\end{table}

%% file: tables/rq1-2-performance.tex
\colorlet{degrade}{red!20}%
\newcommand{\dg}[1]{\cellcolor{degrade!#1}}%
\begin{table}[t]
\caption{\spd\ performances for different data splitting strategies and input contexts. Values in parentheses denote std.\ in the last decimal place over three seeds. }
\centering

\setlength{\tabcolsep}{2pt}
\begin{adjustbox}{width=\columnwidth}
\begin{tabular}{l | l || c | c || c | c || c | c}
\toprule
\multirow{2}{*}{\textbf{Model}} & \multirow{2}{*}{\textbf{Context}} & \multicolumn{2}{c||}{\textbf{Random}} & \multicolumn{2}{c||}{\textbf{Temporal}} & \multicolumn{2}{c}{\textbf{Group-stratified}} \\
& & \textbf{F1 $\uparrow$} & \textbf{PD-S $\downarrow$} & \hspace{5pt}\textbf{F1 $\uparrow$}\hspace{5pt} & \hspace{5pt}\textbf{PD-S $\downarrow$}\hspace{5pt} & \textbf{F1 $\uparrow$} & \textbf{PD-S $\downarrow$} \\
\midrule
\midrule
\multirow{4}{*}{BL} & Commit & \dg{17}\num{0.78}\,$\mathrm{(0)}$ & \dg{49}\num{0.68}\,$\mathrm{(2)}$ & \dg{74}\num{0.45} & \dg{94}\num{0.97} & \dg{34}\num{0.68}\,$\mathrm{(1)}$ & \dg{66}\num{0.79}\,$\mathrm{(7)}$ \\
 & Message & \dg{14}\num{0.79}\,$\mathrm{(0)}$ & \dg{51}\num{0.71}\,$\mathrm{(2)}$ & \dg{63}\num{0.52} & \dg{91}\num{0.95} & \dg{31}\num{0.70}\,$\mathrm{(1)}$ & \dg{66}\num{0.79}\,$\mathrm{(5)}$ \\
 & \diff\textsubscript{c} & \dg{46}\num{0.62}\,$\mathrm{(0)}$ & \dg{83}\num{0.90}\,$\mathrm{(0)}$ & \dg{97}\num{0.32} & \dg{97}\num{0.99} & \dg{71}\num{0.47}\,$\mathrm{(3)}$ & \dg{97}\num{0.98}\,$\mathrm{(1)}$ \\
 & \diff & \dg{46}\num{0.62}\,$\mathrm{(0)}$ & \dg{83}\num{0.90}\,$\mathrm{(1)}$ & \dg{100}\num{0.30} & \dg{97}\num{0.99} & \dg{71}\textbf{\num{0.47}}\,$\mathrm{(3)}$ & \dg{97}\textbf{\num{0.98}}\,$\mathrm{(0)}$ \\
\midrule
\midrule
\multirow{4}{*}{CB} & Commit & \num{0.88}\,$\mathrm{(0)}$ & \num{0.39}\,$\mathrm{(1)}$ & \dg{51}\num{0.58} & \dg{80}\num{0.88} & \dg{20}\num{0.76}\,$\mathrm{(1)}$ & \dg{49}\num{0.69}\,$\mathrm{(4)}$ \\
 & Message & \dg{3}\num{0.86}\,$\mathrm{(0)}$ & \dg{14}\num{0.47}\,$\mathrm{(2)}$ & \dg{49}\num{0.59} & \dg{97}\num{0.98} & \dg{20}\num{0.77}\,$\mathrm{(0)}$ & \dg{46}\num{0.67}\,$\mathrm{(4)}$ \\
 & \diff\textsubscript{c} & \dg{40}\num{0.65}\,$\mathrm{(1)}$ & \dg{74}\num{0.85}\,$\mathrm{(1)}$ & \dg{77}\num{0.43} & \dg{97}\num{0.98} & \dg{60}\num{0.54}\,$\mathrm{(2)}$ & \dg{97}\num{0.98}\,$\mathrm{(2)}$ \\
 & \diff & \dg{40}\num{0.64}\,$\mathrm{(1)}$ & \dg{74}\num{0.85}\,$\mathrm{(0)}$ & \dg{77}\num{0.43} & \dg{97}\num{0.99} & \dg{60}\textbf{\num{0.53}}\,$\mathrm{(3)}$ & \dg{97}\textbf{\num{0.99}}\,$\mathrm{(1)}$ \\
\midrule
\midrule
\multirow{4}{*}{CT5L} & Commit & \num{0.88}\,$\mathrm{(0)}$ & \num{0.39}\,$\mathrm{(0)}$ & \dg{49}\num{0.59} & \dg{86}\num{0.92} & \dg{20}\num{0.76}\,$\mathrm{(5)}$ & \dg{71}\num{0.82}\,$\mathrm{(7)}$ \\
 & Message & \dg{3}\num{0.87}\,$\mathrm{(0)}$ & \dg{6}\num{0.42}\,$\mathrm{(3)}$ & \dg{40}\num{0.65} & \dg{80}\num{0.87} & \dg{20}\num{0.76}\,$\mathrm{(3)}$ & \dg{37}\num{0.62}\,$\mathrm{(3)}$ \\
 & \diff\textsubscript{c} & \dg{40}\num{0.65}\,$\mathrm{(1)}$ & \dg{77}\num{0.86}\,$\mathrm{(1)}$ & \dg{71}\num{0.46} & \dg{94}\num{0.97} & \dg{71}\num{0.47}\,$\mathrm{(8)}$ & \dg{97}\num{0.98}\,$\mathrm{(1)}$ \\
 & \diff & \dg{40}\num{0.64}\,$\mathrm{(1)}$ & \dg{77}\num{0.86}\,$\mathrm{(0)}$ & \dg{77}\num{0.43} & \dg{91}\num{0.95} & \dg{60}\textbf{\num{0.54}}\,$\mathrm{(4)}$ & \dg{94}\textbf{\num{0.97}}\,$\mathrm{(2)}$ \\
\bottomrule
\end{tabular}
\end{adjustbox}
{
\smallskip
{
\begin{flushleft}
BL: TF-IDF + logistic regression baseline, CB: CodeBERT, CT5L: CodeT5-Large. All on the same \num{512}-token inputs.
\diff\textsubscript{c} retains source code comments. \diff\ has comments stripped.
\end{flushleft}
}}
\vspace{-3mm}%
\label{tab:rq1-2-performance}
\end{table}%

%% file: figures/evaluation/temporal-sensitivity.tex
\begin{figure}[b]
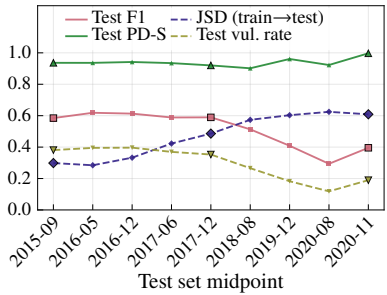

    \centering
    \includesvgwithadjust{.7}{25pt 8pt 0pt 0pt}{data/evaluation/temporal_sensitivity.svg}
    \caption{Sensitivity of CodeBERT trained on \texttt{diff} to temporal split placement. A fixed-size window (20\% train/ val/ test) is shifted across the chronologically ordered dataset in 5\% increments. Outlined markers indicate non-overlapping windows.}%
    \label{fig:temporal-sensitivity}
    \Description{Line chart with four series over nine sliding temporal windows.
    Test F1 (red squares) starts near 0.6 and drops to 0.3 at the 35\% offset
    before partially recovering. Test vulnerability rate (purple triangles)
    follows a similar decline from 0.4 to 0.1. Jensen-Shannon divergence
    (yellow diamonds) rises monotonically from 0.28 to 0.62. PD-S (green
    triangles) stays between 0.75 and 0.99 throughout, showing the model fails
    to detect vulnerabilities at low false-positive rates regardless of window
    placement.}
\end{figure}

%% file: figures/evaluation/model-context-forest.tex
\begin{figure*}[t]
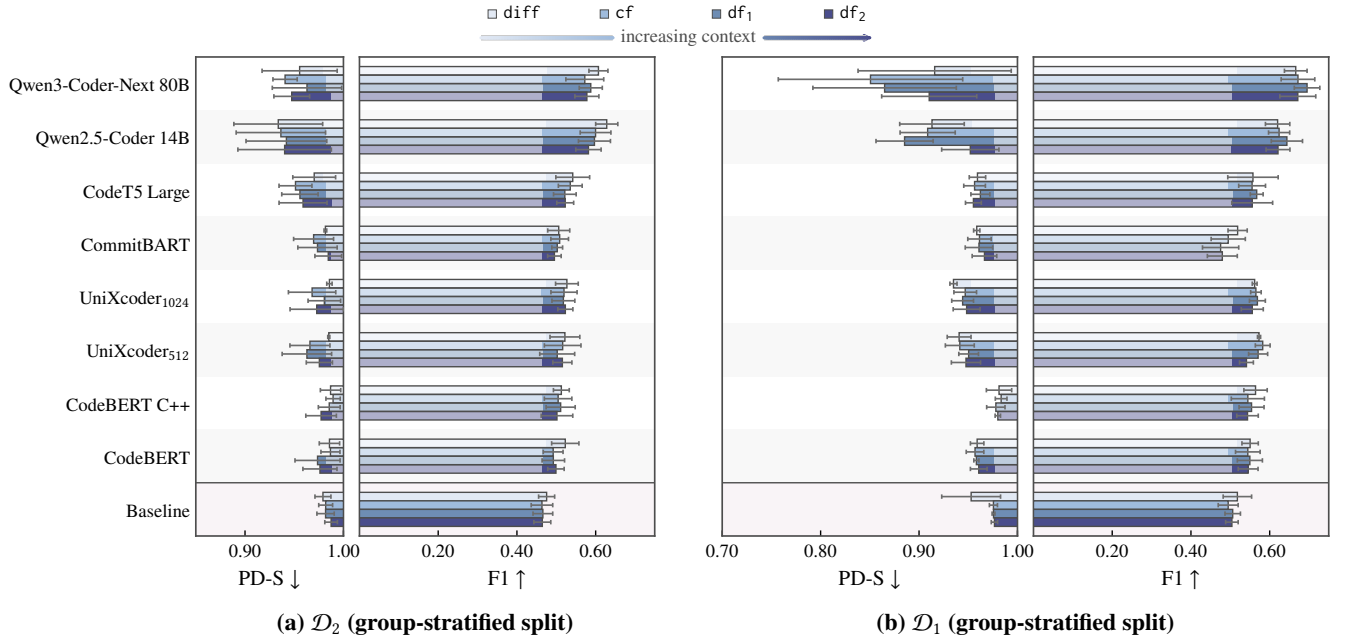

    \centering
    \includesvgwithadjust{1.1}{50pt 0pt 0pt 0pt}{data/evaluation/rq2-model-context-performance.svg}
    \caption{Impact of model capacity and increasing context on \spd\ performance on the group-stratified splits of \ds{1} and \ds{2} over three seeds. Mean alongside standard deviation is reported. Each bar is tinted up to the TF-IDF + LR baseline score on the same representation.}%
    \label{fig:rq2:model-vs-context}
    \Description{Horizontal bar chart with four panels showing F1 and PD-S on D2 and on D1. Rows list seven code language models (UniXcoder at two context lengths) plus a TF-IDF baseline, each with four bars for the diff, cf, df1 and df2 input representations including standard deviation whiskers. Bars are pale up to the score of the TF-IDF baseline on the same representation and fully colored only beyond it. Visible fully colored segments appear mainly for the Qwen models, and no enriched representation consistently outperforms the plain diff.}
\end{figure*}

%% file: tables/repospd-baselines.tex
\begin{table}[b]
\caption{\spd\ performance of baseline approaches under the RepoSPD experimental setup.}
\centering
\setlength{\tabcolsep}{5pt}
\begin{tabular}{l | l || c | c | c | c }
\toprule
\multirow{2}{*}{\textbf{Metric}} & \multirow{2}{*}{\textbf{Model}} & \multicolumn{2}{c|}{\textbf{PatchDB*}} & \textbf{\ds{2}*} & \textbf{\ds{1}*} \\
 & & \textbf{rand} & \textbf{group} & \textbf{group} & \textbf{group} \\
\midrule
\midrule
\multirow{4}{*}{\textbf{F1 $\uparrow$}} & Baseline & \num{0.58} & \num{0.53} & \num{0.51} &  \textbf{\num{0.59}} \\
        & UniXcoder & \num{0.65} & \num{0.62} & \num{0.49} & \textbf{\num{0.59}} \\
        & RepoSPD & \num{0.67} & \num{0.62} & \num{0.53} & \textbf{\num{0.66}} \\
        & Qwen3-C & \num{0.69} & \num{0.67} & \num{0.59} & \textbf{\num{0.68}} \\
\midrule
\multirow{4}{*}{\textbf{PD-S $\downarrow$}} & Baseline & \num{0.94} & \num{0.95} & \num{0.98} & \textbf{\num{0.91}} \\
        & UniXcoder & \num{0.84} & \num{0.88} & \num{0.99} & \textbf{\num{0.93}} \\
        & RepoSPD & \num{0.88} & \num{0.93} & \num{0.99} & \textbf{\num{0.91}} \\
        & Qwen3-C & \num{0.86} & \num{0.95} & \num{0.98} & \textbf{\num{0.82}} \\
\bottomrule
\end{tabular}
\vspace{-3mm}%
\label{tab:repospd-baselines}
\end{table}

%% file: figures/evaluation/attention-attribution.tex
\begin{figure}[t]
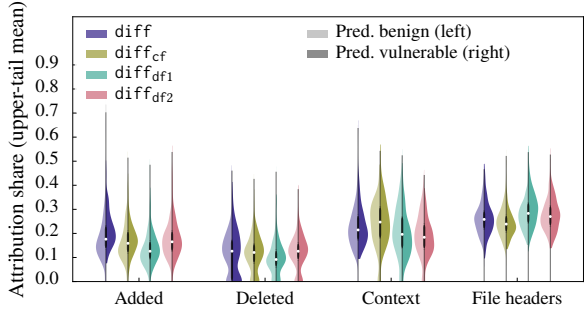

    \centering
      \includesvgwithadjust{0.9}{0pt 0pt 0pt 0pt}{data/evaluation/qwen3-coder-next-base_violin_by_prediction_tail5.svg}
    \caption{Per-sample attribution-share distributions by token category
(average over each category's \num{5}\% most-attributed tokens) using integrated gradients. Qwen3-Coder-Next trained on \ds{1}, evaluated on \num{500} label-balanced commits.}%
    \Description{Violin plots of per-sample integrated gradients attribution shares grouped by token category, including file headers, unchanged context, changed lines, and enriched context statements, across the input representations. The distributions overlap heavily and the changed lines receive no larger attribution share than surrounding context or headers on any representation.}%
    \label{fig:attention-attribution}
\end{figure}

%% file: tables/manual-label-analysis.tex
\begin{table}[ht]
\caption{Manual label analysis: dataset label correctness against adjudicated ground truth per \cve-stratified label group, and the group shares of \ds{1} and \ds{2}. Parenthesized values additionally count unexploitable security-relevant patterns as correct.}
\centering
\setlength{\tabcolsep}{5pt}
\begin{tabular}{l || c || c | c }
\toprule
\multirow{2}{*}{\textbf{Label Group}} & \multirow{2}{*}{\textbf{Corr.\,(\%)}} & \multicolumn{2}{c}{\textbf{Share\,(\%)}} \\
& & \textbf{\ds{1}} & \textbf{\ds{2}} \\
\midrule
\midrule
Benign & \num{92.4} & \num{67.6} & \num{67.6} \\
\vfc\ (non-\cve) & \num{28.8} (\num{66.7}) & \textcolor{gray}{--} & \num{19.5} \\
\vfc\ (\cve) & \num{91.1} & \num{32.4} & \num{13.0} \\
\midrule
Overall & & \num{92.0} & \num{79.8} (\num{87.3}) \\
\bottomrule
\end{tabular}
\vspace{-3mm}%
\label{tab:manual-label-analysis}
\end{table}

%% file: figures/evaluation/error-analysis.tex
\begin{figure}[hb]
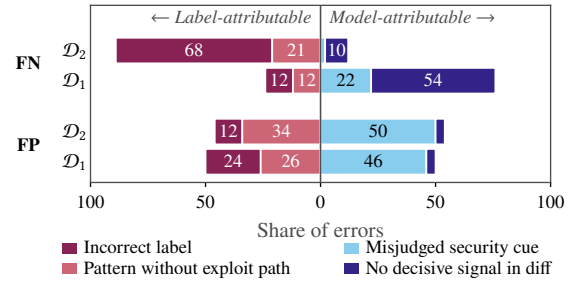

    \centering
    \includesvgwithadjust{1.0}{0pt 0pt 0pt 0pt}{data/evaluation/error_analysis.svg}
    \caption{Manual error analysis of Qwen3-Coder-Next (\diff\ input) on group-stratified splits of \ds{1} and \ds{2}.}%
    \label{fig:error-analysis}
    \Description{Diverging stacked bar chart of error causes for false negatives and false positives on D1 and D2. Each bar splits into four categories that distinguish label-attributable from model-attributable errors. On D2 most false negatives are mislabeled positives, while on D1 the largest false negative category has no decisive signal in the diff. The false positive distributions are similar on both datasets and are dominated by benign changes carrying security-typical patterns.}
\end{figure}

%% file: figures/evaluation/training-dynamics.tex
\begin{figure*}[t]
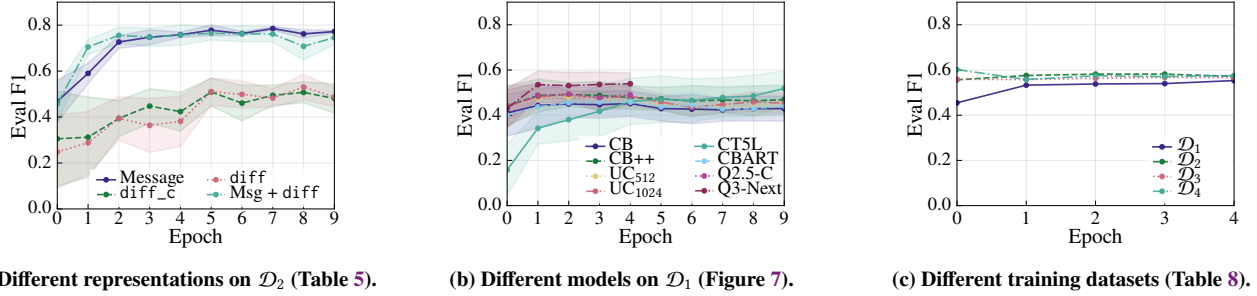

      \centering
          \begin{subfigure}[b]{0.33\linewidth}
      \centering
       \includesvgwithadjust{0.9}{10pt 0pt 10pt 0pt}{data/evaluation/exp1_training_dynamics.svg}
    \caption{Different representations on \ds{2} (Table~\ref{tab:rq1-2-performance}).}%
      \label{fig:eval-representations}
    \end{subfigure}
          \begin{subfigure}[b]{0.33\linewidth}
      \centering
       \includesvgwithadjust{0.9}{10pt 0pt 10pt 0pt}{data/evaluation/exp2_training_dynamics_cveonly.svg}
    \caption{Different models on \ds{1} (Figure~\ref{fig:rq2:model-vs-context}).}%
      \label{fig:eval-models}
    \end{subfigure}
          \begin{subfigure}[b]{0.33\linewidth}
      \centering
       \includesvgwithadjust{0.9}{10pt 0pt 10pt 0pt}{data/evaluation/exp3_training_dynamics.svg}
    \caption{Different training datasets (Table~\ref{tab:cross-dataset-cve}).}%
      \label{fig:eval-datasets}
    \end{subfigure}
    \caption{Training dynamics across evaluation settings: validation F1 per training epoch.}%
    \label{fig:training-dynamics}
    \Description{Training dynamics across different evaluation settings.}
\end{figure*}

%% file: tables/cross-dataset-cve.tex
\begin{table}[b]
\caption{Cross-dataset evaluation: Qwen3-Coder-Next trained on each dataset and evaluated on the \ds{1} validation and test split. All training pools exclude the \ds{1} validation and test groups.}
\centering
\setlength{\tabcolsep}{4pt}
\begin{tabular}{l | c c c c}
\toprule
\textbf{Metric} & \ds{1} & \ds{2} & \ds{3} & \ds{4} \\
\midrule
\textbf{F1 $\uparrow$}  & \num{0.70} & \num{0.70} & \num{0.68} & \num{0.73} \\
\textbf{PD-S $\downarrow$} & \num{0.81} & \num{0.85} & \num{0.90} & \num{0.80} \\
\bottomrule
\end{tabular}
\vspace{-3mm}%
\label{tab:cross-dataset-cve}
\end{table}

%% file: tables/prompt-based.tex
\begin{table}[b]
\caption{Prompt-based \spd\ F1 for GLM-4.7~\citep{GLM45:DBLP:journals/corr/abs-2508-06471} on \ds{1}.}
\centering
\setlength{\tabcolsep}{6pt}
\begin{tabular}{l | c c c c}
\toprule
\textbf{Method} & \textbf{\diff} & \textbf{\texttt{cf}} & \textbf{\texttt{df\textsubscript{1}}} & \textbf{\texttt{df\textsubscript{2}}} \\
\midrule
Direct Answer  & \textbf{\num{0.61}} & \num{0.58} & \num{0.58} & \num{0.58} \\
Chain-of-Thought & \textbf{\num{0.59}} & \num{0.55} & \num{0.55} & \num{0.55} \\
\bottomrule
\end{tabular}
\vspace{-3mm}%
\label{tab:prompt-based}
\end{table}%

%% file: sections/Conclusion.tex
We present a rigorous re-evaluation of code LM-based \spd, consolidating \num{20} datasets and training \num{270} models. The evaluated classifiers exploit natural language signals whenever present, so faithful estimates of code reasoning require code-only inputs and group-stratified splits. Under strict evaluation, no measured configuration reliably detects \vfcs\ from code changes alone. Capacity gains remain insufficient, the evaluated intra- and inter-procedural context enrichments yield no improvements under the strict false positive budget, and a \num{355}B model without fine-tuning only matches far smaller fine-tuned models. Our label audit shows that errors concentrate in \vfcs\ lacking \cve\ association and primarily distort evaluation, while \cve-mapped positives outweigh data volume in our training experiment. The roles of training data, model capacity, and input signal cannot be fully disentangled, but the error analysis suggests the decisive evidence often lies outside the \diff, making the selection and effective use of context fundamental to \spd.

From these findings we derive concrete recommendations for evaluating \spd\ on aggregated \vfc\ corpora. On the protocol side, code understanding should be measured on code-only inputs, generalization estimated over multiple group-stratified splits, and temporal results interpreted in light of compositional shift. On the data side, evaluations should build on \cve-confirmed positives at current label quality, and authors should disclose this narrowed sample distribution and state which notion of a security fix their labels capture and which their system targets.
Building \spd\ systems that are reliable and usable in practice requires selecting and effectively using the context that experts rely on when judging commits, whether through retrieval, codebase specialization, or agentic context gathering. Screening every commit across upstream dependencies keeps efficient specialized classifiers relevant alongside generative approaches. Our framework and evaluation protocol provide a foundation for comparable and rigorous evaluations.

%% file: figures/context-alg.tex
\begin{algorithm}[h!]
  \small 
  \DontPrintSemicolon\
  \SetAlgoLined\
  \KwIn{Pre-patch $c_p$ and target commit $c_t$.}
  \KwOut{Enriched diff $\mathcal{D}_{ctx}$}
  $\mathcal{D}$ $\leftarrow$ diff($c_p$, $c_t$)\;
  $\mathcal{F}$ $\leftarrow$ changed\_functions($\mathcal{D}$)\;
  \For{each $f$ in $\mathcal{F}$}{
    \tcp{Concrete syntax tree}
    $CST_p^f$ $\leftarrow$ TreeSitter($c_p$, $f$)\;
    $CST_t^f$ $\leftarrow$ TreeSitter($c_t$, $f$)\;
    \tcp{Statement level IR}
    ($IR_p^f, IR_t^f$) $\leftarrow$ build\_ir($CST_p^f$, $CST_t^f$)\;

    \tcp{Compute structural diff}
    $\Delta^f$ $\leftarrow$ GumTree($CST_p^f$, $CST_t^f$)\;
    \tcp{Changed statement extraction}
    $(S_\Delta^p, S_\Delta^t)$ $\leftarrow$ extract\_statements($\Delta^f$)\;
    \For{each $k$ in $\{p,t\}$}{
      \tcp{Bi-directional slicing}
      $S_{ctx}^{bw} \leftarrow$ backward\_slice($S_\Delta^k, CST_k^f, d$)\;
      $S_{ctx}^{fw} \leftarrow$ forward\_slice($S_\Delta^k, CST_k^f, d$)\;
      $IR_k^f$ $\leftarrow$ update\_ctx($IR_k^f, S_{ctx}^{bw}, S_{ctx}^{fw}$)\;
    }
    \tcp{Merge and align IRs}
    $IR^f$ $\leftarrow$ merge\_ir($IR_p^f, IR_t^f, \Delta^f$)\;
    $IR^f$ $\leftarrow$ add\_control\_flow($S_\Delta^p$, $S_\Delta^t$)\;
    $\mathcal{D}_{ctx}$ $\leftarrow$ append($\mathcal{D}_{ctx}, IR^f$)\;
  }
  \Return\ $\mathcal{D}_{ctx}$\;
\caption{Context-enriched \texttt{diff}.}%
\label{alg:context-enrichment}%
\end{algorithm}